\documentclass[12pt]{article}
\usepackage{geometry}
\usepackage{rotating}
\usepackage{float}
\usepackage{titling}
\usepackage{authblk}
\usepackage{amssymb}
\usepackage{amsmath}
\usepackage{amsthm}
\allowdisplaybreaks[4]
\usepackage{slashed}
\usepackage{graphicx,color,epsfig}
\usepackage{multirow}
\usepackage{cite}
\usepackage[colorlinks=true,linkcolor=red,citecolor=blue]{hyperref}
\newcommand{\beq}{\begin{eqnarray}}
\newcommand{\eeq}{\end{eqnarray}}

\begin{document}
\title{Systematic Analysis of $B_s \to SP$ Decays in Perturbative QCD Approach}
\author{Zhi-Tian Zou$\footnote{zouzt@ytu.edu.cn}$}
\author{Zhao-Xu He}
\author{Ya-Xin Wang}
\author{Ying Li$\footnote{liying@ytu.edu.cn}$}
\affil{\it Department of Physics, Yantai University, Yantai 264005, China}
\maketitle
\vspace{0.2cm}
\begin{abstract}
Within the perturbative QCD (PQCD) framework, we present a systematic investigation of charmless $B_s \to SP$ decays, where $S$ and $P$ denote scalar and pseudoscalar mesons, respectively. By employing two distinct structural scenarios for scalar mesons, we calculate the branching fractions and direct $CP$ asymmetries for these processes. Our results reveal branching fractions ranging from $10^{-7}$ to $10^{-5}$, values that are well within the measurable range of current experiments. A striking contrast emerges between penguin- and tree-dominated decays: while penguin-dominated processes yield larger branching fractions, tree-dominated decays exhibit significantly enhanced direct $CP$ asymmetries. In particular, the decays $B_s \to f_0(1370) \eta$ and $B_s \to a_0(1450) K$ demonstrate marked sensitivity to the choice of scalar meson scenario, offering critical constraints for identifying the optimal model once experimental data are available. Furthermore, we calculate the branching fractions of $B_s \to f_0(980)(\sigma)P$  decays in two distinct ranges of the mixing angle of $f_0(980)-\sigma$. The dependencies of both branching fractions and $CP$ asymmetries on this mixing angle are rigorously analyzed, establishing a framework essential for determining its value with future experimental results. These findings provide a robust theoretical foundation for advancing the understanding of nonleptonic $B_s$decays in QCD-based formalisms, as well as the nature of scalar mesons.
\end{abstract}

\newpage
\section{Introduction}
The study of $B\to SP$ and $B\to SV$ decays, where $S$, $P$, $V$ stand for scalar, vector, and pseudoscalar mesons, respectively, will provide valuable information on the nature of the scalar mesons. However, it is well known that the identification of scalar mesons is experimentally challenging, and their underlying structure remains theoretically unresolved. In the past few years there are some progresses in the study of charmless hadronic $B$ decays with scalar mesons in the final state both experimentally and theoretically. On the experimental front, numerous measurements of the decay of the $B$ mesons to scalar mesons, such as $f_0(980/1370/1500/1710)$, $a_0(980/1450)$ and $K_0^*(1430)$, have been reported by the Belle \cite{Belle:2004drb, Abe:2005ig, Bondar:2004wr, Belle:2005zev}, BaBar \cite{BaBar:2004eas, BaBar:2005bkm, BaBar:2004pqn, BaBar:2004lds, BaBar:2004bcz, BaBar:2005qms, BaBar:2004mik, BaBar:2005jqu} and LHCb \cite{LHCb:2012qnt, LHCb:2014ioa, LHCb:2015mxr, LHCb:2019vww} collaborations. These results have been summarized and averaged in Refs.~\cite{ParticleDataGroup:2024cfk,HFLAV:2024ctg}. Theoretically, hadronic $B$ decays to scalar mesons have been studied in the QCD-inspired approaches: QCD factorization (QCDF)\cite{Cheng:2005nb, Cheng:2007st, Cheng:2013fba, Cheng:2010sn, Li:2011kw, Chen:2023pms} and perturbative QCD (PQCD) \cite{Wang:2006ria, Shen:2006ms, Zou:2017yxc, Zou:2017iau, Zou:2016yhb, Liu:2021gnp,Zhang:2016qvq, Zhang:2013efa, Zhang:2012zze, Zhang:2010kw, Liu:2009xm}.

In order to study the hadronic charmless B decays containing a scalar meson in the final state, it is necessary to specify the quark content of the scalar meson. It has been suggested that the light scalars below or near 1 GeV--the isoscalars $f_0(600)$ (or $\sigma$), $f_0(980)$, the isodoublet $K_0^*(800)$ (or $\kappa$) and the isovector $a_0(980)$--form an SU(3) flavor nonet, while scalar mesons above 1 GeV, namely, $f_0(1370)$, $a_0(1450)$, $K^*_0(1430)$ and $f_0(1500)/f_0(1710)$, form another nonet. In ref.~\cite{Cheng:2005nb}, Cheng {\it et.al.}explored two possible scenarios: (i) in scenario 1, $\kappa$, $a_0(980)$ and $f_0(980)$ are the lowest lying states, and $K_0^*(1430), a_0(1450), f_0(1500)$ are the corresponding first excited states, respectively, and (ii) in scenario 2 $K_0^*(1430)$, $a_0(1450)$, $f_0(1500)$ are viewed as the lowest lying resonances and the corresponding first excited states lie between $(2.0\sim 2.3)$~GeV. Scenario 2 corresponds to the case that light scalar mesons are four-quark bound states, while all scalar mesons are made of two quarks in scenario 1. The key difference between the two scenarios is how the light scalar mesons, such as $K_0^*(800)$,$a_0(980)$ and $f_0(980)$, are treated, either as the lowest-lying $q\bar{q}$ states or as four-quark particles. Phenomenological studies in Refs. \cite{Cheng:2005nb, Cheng:2007st} suggested that scenario 2 is more favored, a conclusion that is supported by lattice calculations identifying $a_0(1450)$ and $K_0^*(1430)$ as the lowest-lying $S$-wave $q\bar{q}$ states \cite{Mathur:2006bs}. Meanwhile, $a_0(980)$ and $\kappa$ are viewed as $S$-wave tetraquark mesoniums \cite{Prelovsek:2010kg}. However, a competing lattice calculation in ref. \cite{Alexandrou:2012rm} reaches a different conclusion, meaning that the final interpretation remains unresolved. 

In refs. \cite{Zhang:2016qvq, Zhang:2013efa, Zhang:2012zze, Zhang:2010kw, Liu:2009xm}, the authors have studied some of the $B_s \to SP$ decays, however, systematic studies within the PQCD framework remain scarce. In 2019, the LHCb reported the amplitude analysis of $B_s\to K^0_S K^{\pm}\pi^{\mp}$ decays and observed the decays$B_s\to K_0^*(1430)^{\pm}K^{\mp}$ and $B_s\to \overline{K}_0^*(1430)^0K^0/K_0^*(1430)^0\overline{K}^0$ for the first time, each with a significance of over 10 standard deviations. Motivated by these results, we shall comprehensively investigate $B_s \to P S$ decays within the PQCD framework. This work will not only update existing calculations, but also predict decay modes that have not been addressed previously. We will improve upon previous calculations by retaining the terms proportional to $r^2$ (with $r=m_s/m_{B_s}$) in the denominators of the inner quark propagators. This modification significantly alters the behavior of the hard kernel. Furthermore, we will account for the relative sign differences in the vector decay constants between $a_0^-$ and $a_0^+$, as well as between $K_0^*$ and $\overline{K}_0^*$. We also note that in practice it is difficult for us to make quantitative predictions on hadronic $B\to  SP,SV$ decays based on the four-quark picture for light scalar mesons, because their wave functions cannot be defined till now.  Hence, predictions are made only in the 2-quark model for the decays with light scalar mesons. We expect that these systematic and more accurate theoretical calculations will provide valuable insights into the nature of scalar mesons, in conjunction with ongoing precise experimental measurements.

We organize this paper as follows, in Sec.~\ref{sec:2} the framework of PQCD approach and the wave functions of the initial and final states will be introduced,  and the decay amplitudes are also given briefly in this section. In Sec.~\ref{sec:3} we shall present the numerical results of the branching fractions and the direct $CP$ asymmetries of the $B_s\to SP$ decays. Based on the results obtained, some detailed discussions are also performed in this section. Finally, we summarize this work in Sec.~\ref{sec:4}

\section{The Decay Formalism and Wave Functions}\label{sec:2}
The decay rate of two-body $B$-meson hadronic decays cannot be calculated directly from the first principles due to the inherent complexity of QCD. However, PQCD approach, based on $k_T$-factorization, provides a powerful framework for tackling these decays. This method utilizes factorization, allowing the decay amplitude to be expressed as a convolution of contributions from distinct energy scales. For the charmless hadronic two-body $B$-decays, several key energy scales are essential for factorizing the decay amplitude. These include the $W$-boson mass $m_W$, the $b$-quark mass $m_b$ , and the factorization scale $\sqrt{\bar{\Lambda}M_B}$ where $\bar{\Lambda}=M_B-m_b$  represents the difference between thee $B$-meson mass and the $b$-quark mass. At energy scales above the $W$-boson mass, the physics is dominated by electroweak interactions, which can be treated perturbatively. This enables the determination of the Wilson coefficients  $C(m_W)$ at $m_W$ scale. By applying the renormalization group equations, the effects between the $m_W$ and $m_b$ scales are absorbed into the Wilson coefficients $C(m_\mu)$ at a lower scale, which correspond to the relevant four-quark operators. The physics between the $m_b$-scale and the factorization scale is referred to as the “hard kernel” in PQCD. This hard kernel can be computed perturbatively, as it is dominated by the exchange of a single hard gluon. At energy scales below the factorization scale, the physics enters a soft and nonperturbative regime. This is captured by the initial and final hadronic wave functions, which are universal  and encode the nonperturbative effects. Therefore, the decay amplitude is factorized as a convolution of the Wilson coefficients, the hard kernel, and the initial and final hadronic wave functions. 

Thus, we can write the decay amplitude of $B_s\to SP$ decays as
\begin{multline}
 \mathcal{A}=\int dx_1dx_2dx_3b_1db_1b_2db_2b_3db_3 \\
\times {\rm Tr}[C(t)\Phi_{B_s}(x_1,b_1)\Phi_P(x_2,b_2)\Phi_S(x_3,b_3)H(x_i,b_i,t)S_t(x_i)e^{-S(t)}].
\end{multline}
The quantities $x_i$ and $b_i$ represent the longitudinal momentum fractions of the valence quarks and the conjugate space coordinate of the transverse momentum $k_T$ of light quarks in the initial and final mesons, respectively. The scale $t$ corresponds to the largest scale appearing in the hard kernel $H(x_i,b_i,t)$. It is important to emphasize that the key feature of the PQCD approach lies in the inclusion of the transverse momentum $k_T$, which is neglected in collinear factorization schemes such as QCDF. When radiative corrections are considered, the additional scale introduced by the transverse momentum $k_T$ leads to large double logarithms$\ln^2(Q/k_T)$, which can be resummed  to yield the Sudakov factor $e^{-S(t)}$. This factor effectively suppresses soft dynamics and makes the perturbative calculation of the hard kernel applicable at the $m_b$ scale \cite{Li:1997un}. Furthermore, another large double logarithm $\ln^2 x_i$ appears in the radiative corrections, which can also be resummed by threshold resummation to produce the jet function $S_t(x_i)$. This function smears the endpoint singularities at $x_i$ effectively \cite{Li:2001ay}.

The wave functions of both the initial and final states are essential input parameters in the PQCD approach. For the initial $B_s$ meson, its wave function has been extensively studied in $B_s \to PP$, $B_s \to PV$, and $B_s \to VV$ decays \cite{Ali:2007ff, Zou:2015iwa, Qin:2014xta}. The expression for the $B_s$ wave function is given by:
\begin{eqnarray}
\Phi_{B_s}=\frac{1}{\sqrt{2N_c}}(\slashed{P}_{B_s}+M_{B_s})\gamma_5\phi_{B_s}(x,b). \label{eq:wf}
\end{eqnarray}
$\phi_{B_s}(x,b)$ is the light-cone distribution amplitude (LCDA), which is expressed as
\begin{eqnarray}
\phi_{B_s}(x,b)=N_Bx^2(1-x)^2exp\left[-\frac{1}{2}\left(\frac{x M_{B_s}}{\omega}\right)^2-\frac{\omega^2b^2}{2}\right]. \label{eq:lcda}
\end{eqnarray}
In this expression, $N_B$ is a normalization constant that satisfies the normalization condition
\begin{eqnarray}
\int_0^1 dx \phi_{B_s}(x,b=0)=\frac{f_{B_s}}{2\sqrt{6}}.
\end{eqnarray}
The decay constant $f_{B_s}$ is typically taken as $f_{B_s}=(0.23\pm 0.02)~{\rm GeV}$~\cite{Ali:2007ff, Zou:2015iwa, Qin:2014xta, Li:2004ep}, and $\omega$ is the shape parameter, with a typical value of $\omega=(0.5\pm 0.05)~{\rm GeV}$. In current work, we neglect the higher-order terms in $B_s$ meson wave function, because they numerically suppressed  \cite{Li:2004ep, Li:2014xda,Qin:2022rlk}. A model-independent determination of the $B$ meson DA from an Euclidean lattice was attempted in \cite{Wang:2019msf}.

As discussed earlier, there are two prominent scenarios to describe the scalar mesons in the two-quark picture. The definitions of the two types of decay constants for scalar mesons are as follows:
\begin{eqnarray}
\langle S(P)|\bar{q}_2\gamma_{\mu}q_1|0\rangle=f_SP_{\mu} ,\;\;\; \langle S(P)|\bar{q}_2q_1|0\rangle=\bar{f}_Sm_S. \label{eq:fs}
\end{eqnarray}
Here, $m_S$ is the mass of the scalar meson, while the vector decay constant $f_S$  and scalar one $\bar{f}_S$ are related through the equation of motion,
\begin{eqnarray}
\bar{f}_S=\mu_Sf_S,
\end{eqnarray}
where the scale $\mu_S$ is defined as
\begin{eqnarray}
\mu_S=\frac{m_S}{m_2(\mu)-m_1(\mu)}, \label{eq:mus}
\end{eqnarray}
with  $m_1$ and $m_2$ being the masses of the running current quarks in the scalar meson. Due to charge conjugation invariance and the conservation of the vector current, the vector decay constants of some neutral scalar mesons vanish, such as $f_0$, $a_0^0$ and $\sigma$. From eq.~(\ref{eq:mus}), we can directly derive the relationship between the decay constants of a scalar meson and its antiparticle:
\begin{eqnarray}
f_S=-f_{\bar{S}},\;\;\;\bar{f}_S=\bar{f}_{\bar{S}}.
\end{eqnarray}
This relationship is consistent with the definitions of the vector and scalar constants in eq.~(\ref{eq:fs}). For instance, the vector decay constant of $K_0^*$ has the opposite sign to that of $\overline{K}_0^*$.

The wave function of scalar mesons can be expressed in terms of matrix elements involving twist-2 and twist-3 light-cone distribution amplitudes (LCDAs). This wave function is written as:
\begin{eqnarray}
\Phi_S(x)=\frac{i}{\sqrt{6}}\left[P_S\phi_S(x)+m_S\phi_S^s(x)+m_S(\slashed{n}\slashed{v}-1)\phi_S^t(x)\right],
\end{eqnarray}
where $n=(1,0,\mathbf{0}_T)$ and $v=(0,1,\mathbf{0}_T)$ are the light-like unite vector. The functions $\phi_S$ and $\phi_S^{s(t)}$ represent the twist-2 and twist-3 LCDAs. These functions satisfy the following normalization conditions:
\begin{eqnarray}
\int_0^1dx\phi_S(x)=f_S,\;\;\;\int_0^1dx\phi_S^{s(t)}(x)=\bar{f}_S.
\end{eqnarray}
The twist-2 LCDA $\phi_S(x)$ is expanded in terms of Gegenbauer polynomials as \cite{Cheng:2005nb}:
\begin{eqnarray}
\phi_S(x)=\frac{3}{2\sqrt{6}}x(1-x)\left[f_S+\bar{f}_S\sum_{m=1}^{\infty}B_mC_m^{3/2}(2x-1)\right],
\end{eqnarray}
where $C_m^{3/2}(2x-1)$ are the Gegenbauer polynomials and $B_m$ are the corresponding Gegenbauer moments. It is customary to neglect contributions from the even Gegenbauer moments, as these are either zero or small, of order $m_d-m_u$ or $m_s-m_{d,u}$ (see the eq.(C3) in Ref.~\cite{Cheng:2005nb}). Thus, the twist-2 LCDA $\phi_S$  is dominated by the terms associated with the odd Gegenbauer moments, specifically $B_1$ and $B_3$. The values of $B_1$ and $B_3$ have been calculated using the QCD sum rule method in Ref.~\cite{Cheng:2005nb}, and the explicit values can found in Ref.~\cite{Cheng:2013fba}.

For the twist-3 LCDAs, the asymptotic forms are used for simplicity,
\begin{eqnarray}
\phi_S^s(x)=\frac{\bar{f}_S}{2\sqrt{6}},\;\;\;\phi_S^t(x)=\frac{\bar{f}_S}{2\sqrt{6}}(1-2x).
\end{eqnarray}
though the twist-3 LCDAs have also been studied in Refs. \cite{Han:2013zg, Lu:2006fr}. The values of the vector decay constant $f_S$ and the scalar decay constant $\bar{f}_S$  can be found in Ref. \cite{Cheng:2013fba}.

For pseudoscalar mesons, the wave functions have been extensively studied up to the twist-3. They are defined as:
\begin{eqnarray}
\Phi_P(x)=\frac{i}{\sqrt{6}}\left[\slashed{P}\phi_P^A(x)+m_0^P\phi_P^P(x)+m_0^P(\slashed{n}\slashed{v}-1)\phi_P^T(x)\right],
\end{eqnarray}
where the $m_0^P$ is the chiral mass of the pseudoscalar meson, and $\phi_P^{A,P,T}$ are the twist-2 and twist-3 LCDAs. These can also be expanded in terms of Gegenbauer polynomials as
\begin{eqnarray}
\phi_P^A(x)&=&\frac{3f_P}{\sqrt{6}}x(1-x)\left[1+a_1^AC_1^{3/2}(t)+a_2^AC_2^{3/2}(t)+a_4^AC_4^{3/2}(t)\right],\\
\phi_P^P(x)&=&\frac{f_P}{2\sqrt{6}}\left[1+a_2^PC_2^{1/2}(t)+a_4^PC_4^{1/2}(t)\right],\\
\phi_P^T(x)&=&-\frac{f_P}{2\sqrt{6}}\left[C_1^{1/2}(t)+a_3^TC_{3}^{1/2}(t)\right],
\end{eqnarray}
where $t=2x-1$, and the values of the corresponding Gegenbauer moments are taken as
\begin{eqnarray}
&&a_{2\pi}^A=0.44,\;\;\;a_{4\pi}^A=0.25,\;\;\;a_{1K}^A=0.17,\;\;\;a_{2K}^A=0.20,\nonumber\\
&&a_{2\pi}^P=0.43,\;\;\;a_{4\pi}^P=0.09,\;\;\;a_{2K}^P=0.24,\;\;\;a_{4K}^P=-0.11,\nonumber\\
&&a_{3\pi}^T=0.55,\;\;\;a_{3K}^T=0.35.\nonumber
\end{eqnarray}
\begin{figure*}[h]
    \centering
    \includegraphics[width=1.0\linewidth]{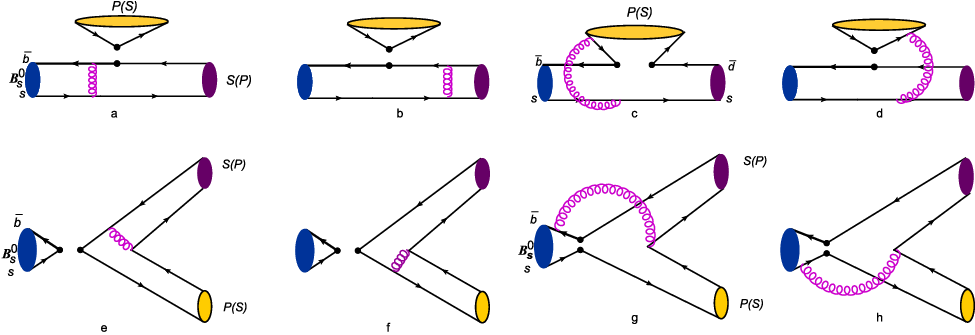}
    \caption{Leading order Feynman diagrams contributing to the $B_{s}\to SP$ decays in POCD approach.} 
    \label{fig:Feynman diagram}
\end{figure*}

We now proceed to calculate the hard kernel $H(x,b,t)$, which is primarily dominated by the exchange of a single hard gluon connecting the four-quark operator with the spectator quark \cite{Lu:2000hj}. The leading-order $B_s\to SP$ decays are described by eight distinct Feynman diagrams, as shown in Fig.\ref{fig:Feynman diagram}. The first row illustrates the emission-type diagrams: the first two are  the factorizable emission diagrams, while the last two are hard-scattering emission diagrams, whose contributions are typically suppressed due to cancellations between them. The second row presents the annihilation-type diagrams. The first two are factorizable, while the last two are non-factorizable. Using the wave functions mentioned earlier, we perform the perturbative calculations within the PQCD framework to obtain the amplitudes for all the diagrams contributing to $B_s\to SP$ decays. By symmetry, the amplitudes for $B_s\to SP$ decays can be obtained by substituting the corresponding relations into the decay amplitudes for $B_s\to SV$ decays, as detailed in Ref. \cite{Liu:2021gnp}.
\begin{eqnarray}
r_V\to r_{0p}, \;\phi_V(x)\to \phi_P^A(x),\;\phi_{V}^S(x)\to\phi_P^P(x),\;\phi_V^t(x)\to\phi_P^T(x),\;f_V\to f_P,
\end{eqnarray}
where $r_{0p}=m_0^P/m_{B_s}$. For the sake of simplicity, we do not repeat the decay amplitudes in this work.

\section{Numerical Results and Discussions}\label{sec:3}
Using the analytic decay amplitudes, we calculate the branching fractions and direct $CP$ asymmetries for all $B_s\to SP$ decays within the two-quark model of scalar mesons. As aforementioned,  although the four-quark picture is better suited for explaining certain phenomena—such as the mass degeneracy of $f_0(980)$ and $a_0(980)$, and the broader widths of $K_0(800)$ and $\sigma(500)$ —the study of it extends beyond the current theoretical framework. In Tables \ref{table:1} and \ref{table:2}, we present the branching fractions and direct $CP$ symmetries for $B_s\to PS$ decays involving scalar mesons with masses below 1 GeV. Predictions for the branching fractions and direct $CP$ asymmetries for decays to heavier scalar mesons (around 1.5 GeV) in scenario 1 and scenario 2   are given in Tables \ref{table:3} and \ref{table:4}, respectively. 


In this study, we consider three sources of error to account for theoretical uncertainties in our perturbative calculations. The first source arises from uncertainties in the hard parameters within the wave functions of the  $B_s$ meson, pseudoscalar mesons, and scalar mesons. These include decay constants ($f_{B_s}$, $f_S$, $\bar{f}_S$ and $f_P$), the shape parameter $\omega$ in the LCDA of the $B_s$ meson, and the Gegenbauer moments ($B_1$, $B_3$ and $a_i$) in the LCDAs of the final mesons. These uncertainties represent the dominant source of error, particularly for the branching fractions, since the hadronic wave functions of both the initial and final mesons are crucial non-perturbative inputs in the PQCD approach. It is important to note that this source of error will diminish as experimental data and non-perturbative models improve. However, direct $CP$ asymmetries are relatively insensitive to these uncertainties, as the errors from the wave functions tend to cancel out in the ratio, which is defined as a fraction. The second source of error pertains to uncertainties arising from higher-order radiative corrections and power corrections. These are accounted for by varying the QCD scale $\Lambda_{QCD}=0.25\pm0.05~{\rm GeV}$ and the hard scale $t$, where we vary $t$ between $0.8t$ and $1.2 t$. As indicated in the tables, the direct $CP$ asymmetries are sensitive to these errors, as higher-order corrections significantly affect the strange quark phase, which plays a critical role in determining the $CP$ asymmetry. Finally, we assess the uncertainties stemming from the CKM matrix elements.

\begin{table}[!t]
    \caption{The $CP$-averaged branching fractions ($\cal B$)  and the direct $CP$ asymmetries ($\mathcal{A}_{CP}^{dir}$) of the $B_s\to a_0(980)(\kappa)P$ decays in scenario 1.}
     \label{table:1}
    \begin{center}
    \begin{tabular}{l |c |c }
     \hline \hline
    Decay Modes&${\cal B}(10^{-6})$&$\mathcal{A}_{CP}^{dir}$($\%$)   \\
    \hline\hline
     $B_s \to a_0\pi^0$
     &$0.37^{+0.18+0.17+0.09}_{-0.09-0.03-0.00}$
     &$10.5^{+1.4+0.4+0.6}_{-0.9-0.2-0.7}$ \\

     $B_s \to a_0\eta$
     &$0.04^{+0.01+0.01+0.00}_{-0.01-0.01-0.00}$
     &$-2.30^{+0.40+0.39+0.11}_{-0.28-2.45-0.00}$ \\
     
     $B_s \to a_0\eta^{\prime}$
     &$0.09^{+0.02+0.01+0.00}_{-0.02-0.04-0.01}$
     &$-4.80^{+2.85+1.35+0.16}_{-0.18-1.39-0.55}$    \\

     $B_s \to a_0^+K^{-}$
     &$1.31^{+0.37+0.39+0.09}_{-0.19-0.27-0.02}$
     &$97.4^{+0.9+0.9+1.0}_{-9.10-0.5-1.3}$  \\

     $B_s \to a_0\overline{K}^{0}$
     &$2.05^{+1.05+0.26+0.11}_{-0.74-0.37-0.08}$
     &$71.6^{+8.0+7.5+0.5}_{-10.2-6.3-0.5}$   \\

     $B_s \to a_0^+\pi^-$
     &$0.44^{+0.12+0.08+0.01}_{-0.11-0.12-0.02}$
     &$-19.5^{+4.4+2.9+1.3}_{-1.6-2.2-1.5}$   \\

     $B_s \to a_0^-\pi^+$
     &$0.37^{+0.13+0.05+0.01}_{-0.11-0.08+0.01}$
     &$42.8^{+0.5+2.1+2.7}_{-3.5-1.4-2.4}$ \\

     $B_s \to \kappa^- K^{+}$
     &$5.55^{+3.21+2.23+0.06}_{-1.72-1.77-0.11}$
     &$-42.6^{+8.0+4.3+2.0}_{-5.0-3.7-2.4}$ \\

     $B_s \to \kappa^+K^{-}$
     &$11.4^{+3.0+3.5+0.3}_{-2.9-3.7-0.5}$
     &$-6.96^{+2.14+1.21+0.47}_{-1.80-0.89-0.55}$  \\

     $B_s \to \kappa^0 \overline{K}^{0}$
     &$11.6^{+0.37+0.33+0.03}_{-0.32-0.35-0.05}$
     & 0.0\\

     $B_s \to \overline \kappa^0 K^{0}$
     &$5.92^{+2.05+2.37+0.24}_{-1.79-1.76-0.17}$
     & 0.0 \\

     $B_s \to \kappa^- \pi^+$
     &$7.85^{+1.60+0.59+0.44}_{-1.52-0.88-0.35}$
     &$18.4^{+2.8+3.8+0.0}_{-3.2-3.4-0.1}$    \\

     $B_s \to \overline \kappa^0\pi^0$
     &$0.15^{+0.06+0.03+0.01}_{-0.04-0.02-0.00}$
     &$36.5^{+25.1-27.0-0.0}_{-20.4-16.7-6.3}$   \\

     $B_s \to  \overline \kappa^0 \eta$
     &$0.15^{+0.03+0.17+0.00}_{-0.02-0.05-0.02}$  &$65.0^{+4.5+4.7+0.7}_{-4.8-2.2-1.2}$    \\

     $B_s \to \overline \kappa^0 \eta^{\prime}$
     &$0.56^{+0.13+0.15+0.00}_{-0.15-0.19-0.07}$
     &$-27.0^{+2.4+3.5+0.0}_{-2.8-6.6-1.5}$\\
     \hline \hline
    \end{tabular}
    \end{center} 
    \end{table}

    \begin{table}[!t]
     \caption{The $CP$-averaged branching fractions ($\cal B$)  and the direct $CP$ asymmetries ($\mathcal{A}_{CP}^{dir}$) of the $B_s\to \sigma P$ and $B_s\to f_0(980) P$ with the mixing angle $\theta$ in scenario 1.}    \label{table:2}
     \centering
     \begin{tabular}[t]{l |c| c| c| c}
    \hline\hline
      \multirow{2}{*}{Decay Modes} & \multicolumn{2}{|c|}{$[25^{\circ}, 40^{\circ}]$} & \multicolumn{2}{|c}{$[140^{\circ}, 165^{\circ}]$} 
    \\\cline{2-5}
      &${\cal B}(10^{-6})$&$A_{CP}(\%)$&${\cal B}(10^{-6})$&$A_{CP}(\%)$\\
     \hline 
    $B_s\to \sigma\overline{K}^{0}$ &$0.83\sim1.20$&$-35.2\sim-3.05$&$1.79\sim1.95$&$-71.6\sim-69.3$\\
    $B_s\to \sigma \pi^{0}$ &$0.016\sim0.028$ &$3.33\sim20.1$&$0.014\sim0.038$&$38.3\sim43.6$\\  
    $B_s\to \sigma\eta$ &$1.69\sim4.58$ &$-18.3\sim-8.75$&$5.93\sim12.0$&$2.55\sim3.51$\\  
    $B_s\to \sigma\eta^{\prime}$ &$1.61\sim5.81$ &$-26.6\sim-10.5$&$16.2\sim28.7$&$2.96\sim3.81$\\   
    $B_s\to f_0(980)\overline{K}^{0}$ &$1.23\sim1.58$&$-75.3\sim-52.1$&$0.49\sim0.62$&$14.0\sim47.7$\\
    $B_s\to f_0(980)\pi^{0}$ &$0.085\sim0.11$ &$19.9\sim22.3$&$0.073\sim0.12$&$8.52\sim14.0$\\   
    $B_s\to f_0(980)\eta$ &$17.0\sim20.9$ &$0.63\sim1.60$&$10.0\sim18.7$&$-5.56\sim-2.26$\\  
    $B_s\to f_0(980)\eta^{\prime}$ &$26.1\sim31.3$ &$4.61\sim4.84$&$15.5\sim25.9$&$4.69\sim5.01$\\
    \hline\hline
    \end{tabular}
    \end{table}
    
\begin{sidewaystable}
    \caption{The $CP$ averaged branching fractions ($\cal B$) and the direct $CP$ asymmetries ($\mathcal{A}_{CP}^{dir}$) of the $B_s\to S P$ decays involving the heavier scalar mesons in scenario 1.} \label{table:3}
    \begin{center}
     \begin{tabular}[t]{l |c| c| c| c}
    \hline\hline
      \multirow{2}{*}{Decay Modes} & \multicolumn{2}{|c|}{${\cal B}(10^{-6})$} & \multicolumn{2}{|c}{$\mathcal{A}_{CP}^{dir}(\%)$} 
    \\\cline{2-5}
      &Current Work &Former Results&Current Work &Former Results\\
     \hline 
    \hline

     $B_s \to a_0(1450)\pi^0$
     &$2.41^{+0.82+0.70+0.00}_{-1.31-0.98-0.03}$
     &
     &$12.2^{+1.5+1.2+1.0}_{-1.3-0.7-0.7}$       
     &
     \\              

     $B_s \to a_0(1450)\eta$
     &$0.03^{+0.01+0.01+0.00}_{-0.02-0.01-0.00}$
     &      
     & $-5.25^{+0.80+1.70+0.54}_{-1.34-1.60+0.22}$  
     &          
     \\
     
     $B_s \to a_0^+(1450)K^{-}$
     &$1.48^{+0.55+0.23+0.04}_{-0.69-0.34-0.09}$
     &   
     &$3.28^{+0.69+0.63+0.01}_{-1.01-0.40-0.01}$  
     &          
     \\ 

     $B_s \to a_0(1450)\bar{K}^{0}$
     &$2.26^{+0.89+0.48+0.08}_{-0.95-0.51-0.12}$
     & 
     &$-7.47^{+2.30+2.00+0.06}_{-1.91-1.63-0.08}$     
     &    
     \\

     $B_s \to a_0^+(1450)\pi^-$
     &$2.55^{+1.03+0.08+0.08}_{-0.90-0.31-0.09}$
     & 
     &$-5.56^{+1.09+0.08+0.59}_{-1.68-0.59-0.67}$     
     &    
     \\

     $B_s \to a_0^-(1450)\pi^+$
     &$2.49^{+0.94+0.27+0.01}_{-0.86-0.34-0.09}$
     &   
     &$28.6^{+2.0+2.3+0.9}_{-2.1-2.7-0.7}$  
     &   
     \\

     $B_s \to a_0(1450)\eta^{\prime}$
     &$0.08^{+0.03+0.02+0.00}_{-0.03-0.02-0.00}$
     &   
     &$-4.69^{+1.33+0.83+0.50}_{-1.17-0.75-0.63}$ 
     &   
     \\

     $B_s \to K_0^{*-}(1430) K^{+}$
     &$24.6^{7.1+5.7+0.7}_{-6.0-4.3-0.1}$
     &   
     &$-10.5^{+2.2+2.8+1.3}_{-1.4-5.0-0.3}$ 
     &   
     \\

     $B_s \to K_0^{*+}(1430)K^{-}$
     &$19.0^{+8.6+5.1+2.8}_{-6.5-4.4-1.0}$
     &   
     &$-1.32^{+0.53+2.18+0.00}_{-0.43-2.05-0.14}$ 
     &   
     \\

     $B_s \to K_0^{*0}(1430)\bar{K}^{0}$
     &$21.5^{+6.7+5.3+0.3}_{-7.5-4.8-1.2}$
     &    
     &$0.0$ 
     &   
     \\
          
     $B_s \to \bar{K}^{*0}(1430)K^{0}$
     &$21.9^{+7.9+6.0+0.4}_{-6.7-7.0-0.9}$
     &    
     &$0.0$ 
     &   
     \\

     $B_s \to K_0^{*-}(1430)\pi^+$
     &$6.90^{+2.9+1.3+0.3}_{-2.5-1.2-0.3}$
     &$12^{+5+2+1}_{-3-2-1}$  
     &$34.9^{+5.3+4.2+1.0}_{-4.8-3.7-0.0}$   
     &$28.1^{+4.9+0.7+0.0}_{-4.4-0.7-2.4}$
     \\     

      $B_s \to \bar{K}^{*0}(1430)\eta$
      &$0.30^{+0.16+0.06+0.02}_{-0.12-0.05-0.02}$   
      &$0.385^{+0.148+0.081+0.041+0.091}_{-0.101-0.077-0.037-0.06}$    
      &$70.4^{+4.9+16.7+2.3}_{-5.8-20.5-2.4}$  
      &$42.1^{+6.9+0.0+2.8+14.2}_{-5.9-0.0-3.3-12.1}$     
      \\ 

      $B_s \to \bar{K}^{*0}(1430)\pi^0$
      &$0.35^{+0.19+0.07+0.01}_{-0.14-0.07-0.00}$
      &$0.45^{+0.14+0.09+0.04}_{-0.11-0.09-0.04}$        
      &$95.0^{+0.8+3.1+0.6}_{-2.8-6.0-1.5}$  
      &$93.6^{+4.8+1.0+0.0}_{-7.3-1.0-8.7}$     
       \\
       
      $B_s \to \bar{K}^{*0}(1430)\eta^{\prime}$  
      &$1.09^{+0.50+0.40+0.01}_{-0.39-0.14-0.0.00}$  
      &$0.52^{+0.16+0.11+0.06+0.08}_{-0.10-0.01-0.06-0.08}$    
      &$-41.5^{+2.9+14.0+1.3}_{-3.4-14.1-1.4}$   
      &$70.9^{+2.8+0.0+7.6+17.0}_{-6.0-0.0-2.9-16.5}$
      \\

      $B_s \to f_0(1370)\bar{K}^{0}$     
      &$0.84^{+0.51+0.20+0.03}_{-0.41-0.16-0.01}$    
      &    
      &$-39.3^{+8.5+2.9+0.4}_{-2.7-1.8-0.3}$ 
      &  
      \\

      $B_s\to f_0(1370)\pi^0$         
      & $0.04^{+0.02+0.01+0.00}_{-0.02-0.01-0.00}$ 
      &    
      &$-47.9^{+5.9+12.2+4.1}_{-6.8-15.4-6.2}$ 
      &  
      \\
 
      $B_s\to f_0(1370)\eta$   
      &  $1.25^{+0.69+0.64+0.05}_{-0.54-0.48-0.08}$  
      &    
      &$12.1^{+2.1+3.1+1.1}_{-1.9-1.7-1.5}$ 
      &  
      \\

      $B_s\to f_0(1370)\eta^{\prime}$          
      & $16.1^{+9.4+6.2+0.4}_{-8.3-5.0-0.4}$    
      &    
      &$10.9^{+3.1+2.6+0.7}_{-4.2-2.1-0.5}$ 
      &  
      \\

      $B_s\to f_0(1500)\bar{K}^{0}$   
      & $0.41^{+0.24+0.13+0.02}_{-0.18-0.07-0.01}$ 
      &$0.97^{}_{}$   
      &$64.7^{+4.5+4.8+0.7}_{-14.0-3.0-1.1}$  
      &$47.4^{}_{}$   
      \\ 

      $B_s\to f_0(1500)\pi^0$          
      &$0.05^{+0.03+0.01+0.01}_{-0.02-0.01-0.00}$  
      &$(4.46^{+0.11+2.47+0.68}_{-0.10-1.85-0.63})\times 10^{-2}$
      & $14.3^{+7.3+0.6+1.7}_{-7.5-1.7-0.9}$  
      &$-27.5^{+0.0+0.2+4.1}_{-0.0-0.0-3.3}$   
      \\

      $B_s\to f_0(1500)\eta$         
      &$25.4^{+16.4+9.8+0.6}_{-13.1-7.9-0.6}$  
      & 
      & $-3.23^{+1.33+1.21+1.07}_{-1.41-0.59-0.09}$  
      &  
      \\  

      $B_s\to f_0(1500)\eta^{\prime}$  
      & $33.3^{+20.4+13.0+0.7}_{-16.7-10.2-0.9}$  
      &  
      &$-2.43^{+0.42+0.16+0.14}_{-1.16-0.15-0.09}$  
      & 
      \\  
     \hline \hline
    \end{tabular}
    \end{center}
\end{sidewaystable}

\begin{sidewaystable}
    \caption{The $CP$ averaged branching fractions ($\cal B$) and the direct $CP$ asymmetries ($\mathcal{A}_{CP}^{dir}$)  of the $B_s\to S P$ decays involving the heavier scalar mesons in scenario 2.}. \label{table:4}
    \begin{center}
     \begin{tabular}[t]{l |c| c| c| c}
    \hline\hline
      \multirow{2}{*}{Decay Modes} & \multicolumn{2}{|c|}{${\cal B}(10^{-6})$} & \multicolumn{2}{|c}{$\mathcal{A}_{CP}^{dir}(\%)$} 
    \\\cline{2-5}
      &Current Work &Former Results&Current Work &Former Results\\
     \hline 
    \hline

     $B_s \to a_0(1450)\pi^0$
     &$1.57^{+0.94+0.96+0.06}_{-0.69-0.76-0.00}$
     &  
     &$12.3^{+0.5+0.1+0.0}_{-0.4-0.8-0.9}$    
      &
     \\

     $B_s \to a_0(1450)\eta$
     &$0.01^{+0.01+0.00+0.00}_{-0.01-0.00-0.00}$
     &  
     & $-4.43^{1.73+2.52+0.21}_{-0.12-0.90-0.14}$   
     & 
     \\

     $B_s \to a_0^+(1450)K^{-}$
     &$2.08^{+0.40+0.50+0.00}_{-0.84-0.65-0.15}$
     & 
     &$72.5^{+4.3+1.4+0.1}_{-3.2-1.8-0.1}$    
     &
     \\

     $B_s \to a_0(1450)\bar{K}^{0}$
     &$1.61^{+0.83+0.21+0.13}_{-0.65-0.02-0.09}$
     &  
     &$94.2^{+11.8+15.3+2.19}_{-12.3-14.7-1.77}$    
     &
     \\

     $B_s \to a_0^+(1450)\pi^-$
     &$1.82^{+0.83+0.21+0.05}_{-0.67-0.19-0.06}$
     &  
     &$-2.31^{+0.67+0.13+0.15}_{-0.71-0.09-0.13}$     
     &
     \\

     $B_s \to a_0^-(1450)\pi^+$
     &$1.51^{+0.78++0.14+0.05}_{-0.64-0.21-0.07}$
     &   
     &$26.4^{+1.2+0.9+1.1}_{-1.1-1.2-1.2}$    
     &
     \\

     $B_s \to a_0(1450)\eta^{\prime}$
     &$0.04^{+0.02+0.01+0.00}_{-0.02-0.01-0.00}$
     &
     &$-3.22^{+0.55+0.63+0.03}_{-0.80-0.97-0.02}$     
     &
     \\

     $B_s \to K_0^{*-}(1430) K^{+}$
     &$21.6^{+6.8+3.4+0.7}_{-6.6-3.2-1.3}$
     &  
     &$-41.8^{+9.3+5.5+2.7}_{-14.6-4.9-3.7}$   
     & 
     \\

     $B_s \to K_0^{*+}(1430)K^{-}$
     &$33.3^{+16.1+10.7+1.3}_{-12.5-7.4-1.0}$
     &  
     &$-3.22^{+0.40+0.49+0.07}_{-0.46-0.36-0.08}$   
     &  
     \\

     $B_s \to K_0^{*0}(1430)\bar{K}^{0}$
     &$33.3^{+16.3+4.1+0.6}_{-12.8-7.3-1.6}$
     &  
     $0.0$     
     &
     \\

     $B_s \to \bar{K}^{*0}(1430)K^{0}$
     &$20.7^{10.4+3.5+0.4}_{-8.8-3.3-0.9}$
     &  
     $0.0$     
     &
     \\

     $B_s \to K_0^{*-}(1430)\pi^+$
     &$22.7^{+7.2+1.2+1.0}_{-6.5-1.7-0.7}$
     &$37^{+14+9+8}_{-10-8-7}$
     &$22.8^{+3.5+4.1+1.6}_{-5.1-0.9-0.0}$    
     &$21.0^{+3.4+2.1+0.0}_{-3.1-2.5-2.5}$ 
     \\
 
      $B_s \to \bar{K}^{*0}(1430)\eta$
      &$0.54^{+0.36+0.14+0.03}_{-0.29-0.11-0.03}$  
      &$0.387^{+0.122+0.089+0.282+0.084}_{-0.083-0.080-0.173-0.090}$  
      &$61.1^{+3.1+7.1+1.8}_{-3.3-5.7-2.0}$   
      &$56.2^{+1.2+0.0+7.6+17.0}_{-2.0-0.1-5.9-15.7}$ 
      \\

      $B_s \to \bar{K}^{*0}(1430)\pi^0$
      &$0.19^{+0.13+0.06+0.01}_{-0.11-0.04-0.01}$  
      &$0.41^{+0.10+0.10+0.06}_{-0.07-0.01-0.06}$ 
      &$83.3^{+9.0+9.2+3.1}_{-2.6-2.4-0.8}$  
      &$95.5^{+1.2+3.4+4.0}_{-8.7-4.1-13.9}$ 
     \\

      $B_s \to \bar{K}^{*0}(1430)\eta^{\prime}$
      &$1.53^{+0.82+0.32+0.01}_{-0.64-0.23-0.01}$
      &$0.78^{+0.21+0.18+0.31+0.12}_{-0.13-0.16-0.27-0.13}$ 
      &$-13.5^{+0.7+1.6+0.4}_{-0.7-1.4-0.5}$  
      &$42.4^{+1.1+0.1+7.0+8.5}_{-3.1-0.1-5.9-10.1}$
      \\

      $B_s \to f_0(1370)\bar{K}^{0}$         
      &$1.60^{+0.95+0.32+0.05}_{-0.78-0.35-0.00}$  
      &  
      &$-78.7^{+15.8+4.3+2.3}_{-19.8-6.6-2.7}$    
      &
      \\

      $B_s\to f_0(1370)\pi^0$        
      &$0.11^{+0.06+0.01+0.01}_{-0.05-0.01-0.01}$  
      &  
      &$-8.03^{+1.47+3.55+0.22}_{-2.13-3.72-0.09}$  
      &
      \\

      $B_s\to f_0(1370)\eta$       
      &$14.3^{+9.0+3.1+0.4}_{-6.4-2.8-0.3}$  
      & 
      &$-10.3^{+4.7+3.7+0.3}_{+2.1+2.4+0.4}$  
      &
      \\

      $B_s\to f_0(1370)\eta^{\prime}$      
      &$28.9^{+15.2+6.2+0.8}_{-13.0-5.6-0.64}$  
      & 
      &$2.99^{+2.31+1.13+0.64}_{-2.17-0.86-0.25}$  
      &
      \\

      $B_s\to f_0(1500)\bar{K}^{0}$   
      &$0.37^{+0.18+0.07+0.01}_{-0.15-0.05-0.00}$ 
      &$0.95^{}_{}$ 
      &$32.8^{+6.6+1.8+1.0}_{-8.3-2.7-1.2}$ 
      &$62.8^{}_{}$ 
      \\

      $B_s\to f_0(1500)\pi^0$     
      &$0.23^{+0.06+0.02+0.01}_{-0.06-0.04-0.01}$   
      &$0.28^{+0.06+0.03+0.05}_{-0.05-0.03-0.04}$ 
      &$14.4^{+3.7+1.2+1.4}_{-5.4-1.2-1.2}$  
      &$-9.7^{+0.0+1.2+5.8}_{-0.0-1.1-4.9}$
      \\

      $B_s\to f_0(1500)\eta$    
      &$37.7^{+23.7+8.1+1.0}_{-17.0-7.04-0.8}$  
      & 
      $-1.46^{+0.41+0.43+0.05}_{-0.95-0.20-0.00}$      
      &
      \\

      $B_s\to f_0(1500)\eta^{\prime}$  
      &$43.1^{+26.2+9.9+1.1}_{-18.6-8.7-1.4}$  
      &  
      &$-1.01^{+0.83+0.25+0.15}_{-0.69-0.13-0.07}$  
      &   
      \\
     
     \hline \hline
    \end{tabular}
    \end{center}
 \end{sidewaystable}
 
In Table.~\ref{table:1}, we present the branching fractions for the decays $B_s\to a_0(980) P$ and $B_s\to \kappa P$, which are calculated for the first time using PQCD. These decays are primarily driven by the $\bar{b} \to \bar{s}$ transition, and their rates are significantly enhanced by the large CKM elements $|V_{tb}V_{ts}|$. As a result, the branching fractions for $B_s\to  \kappa K$ can reach values as large as $10^{-6}$, or even higher. In contrast, the branching fractions for $B_s\to a_0(980)\pi(\eta^{(\prime)})$, although governed by the same $\bar{b}\to\bar{s}$ transition and CKM factors, are much smaller—typically on the order of $10^{-7}$ or less. This suppression occurs because these decays are either purely annihilation-type or dominated by annihilation-type contributions, which are power-suppressed.  For the decays $B_s\to \rho^0\overline{K}^{*0}$, $B_s\to \pi^0\overline {K}^{*0}$, $B_s\to \rho^0\overline{K}^{0}$, and $B_s\to \pi^0\overline{K}^{0}$, the branching fractions are significantly smaller than those for corresponding decays with charged mesons in the final state, such as  $B_s\to \rho^+K^{*-}$, $B_s\to \pi^+ K^{*-}$, $B_s\to \rho^+K^{-}$, and $B_s\to \pi^+K^{-}$ \cite{Ali:2007ff, Zou:2015iwa}. This difference arises due to the large color-allowed emission-type diagrams, which are enhanced by the large CKM elements $|V_{ub}V_{ud}|$ in decays involving charged final states. Similarly, the branching fraction for $B_s\to \overline\kappa^0\pi^0$  is much smaller than for the corresponding decay with charged mesons, $B_s\to \kappa^-\pi^{+}$, reflecting the same trend. However, for the decays $B_s\to a_0^+(980)K^-$ and $B_s\to a_0^0(980) \overline{K}^0$, the situation reverses. The branching fraction for $B_s\to a_0^+(980)K^-$ is approximately half that of $B_s\to a_0^0(980)\overline{K}^0$, which is consistent with the trend observed in the decays $B_s\to a_0^+(980)K^{*-}$ and $B_s\to a_0^0(980) \overline{K}^{*0}$ \cite{Liu:2021gnp}. This difference arises because the large emission-type diagrams are highly suppressed in $B_s\to a_0^+(980)K^-$ and $B_s\to a_0^+(980)K^{*-}$  due to the small vector decay constant of $a_0^+(980)$. On the other hand, the decays $B_s\to a_0^0(980) \overline{K}^0$ and $B_s\to a_0^0(980) \overline{K}^{*0}$ are enhanced by hard-scattering emission diagrams, with a large Wilson coefficient $C_2$ and the emission of the $a_0^0(980)$ meson. It is also important to note that, as is typical, hard-scattering emission diagrams involving a pseudoscalar meson and a vector meson are suppressed, since the two diagrams cancel each other due to symmetry. However, when a scalar meson is emitted, the two hard-scattering emission diagrams do not cancel; instead, they amplify each other. In addition, we also obtain the ratio as
  \begin{eqnarray}
R=\frac{{\cal B}(B_s\to a_0^+(980)K^-)}{{\cal B}(B_s\to a_0^0(980)\overline{K}^0)}
 =\frac{1.31}{2.05}\approx\frac{{\cal B}(B_s\to a_0^+(980)K^{*-})}{{\cal B}(B_s\to a_0^0(980)\overline{K}^{*0})}=\frac{2.06}{4.09}\approx\frac{1}{2},
  \end{eqnarray}
which can used to test the application of PQCD.

For decay modes dominated by penguin diagrams, such as $B_s\to a_0(980)\eta^{(\prime)}$ and $B_s\to \kappa^+K^-$, the contributions from tree-level operators are heavily suppressed. This suppression is due to both the small CKM matrix elements $|V_{us}V_{ub}|$ and the small vector decay constants of the scalar mesons. Consequently, the direct $CP$ asymmetries for these decays are very small, since the direct $CP$ asymmetry is proportional to the interference between tree-level and penguin contributions. For the pure annihilation decay $B_s\to a_0^0(980)\pi^0$, the direct $CP$ asymmetry is notably smaller than those for the decays $B_s\to a_0^+(980)\pi^-$ and $B_s\to a_0^-(980)\pi^+$, which are also pure annihilation processes. This occurs because, in $B_s\to a_0^0(980)\pi^0$, the tree-level contribution is strongly suppressed due to a cancellation between $W$-exchange annihilation diagrams and the corresponding particle-exchange diagrams.In contrast, for the decays $B_s\to a_0^+(980)\pi^-$ and $B_s\to a_0^-(980)\pi^+$, there is only a single large tree-level contribution arising from $W$-exchange annihilation diagrams. 

For tree-dominated decays, such as $B_s\to \kappa \pi$, the direct $CP$ asymmetries are significantly larger than those for $B_s\to a_0(980)K$. For the decay $B_s\to \kappa^-\pi^+$, where the $\pi^+$ meson is emitted, the emission diagrams are enhanced by both large Wilson coefficients $C_1/3+C_2$ and large CKM elements ($|V_{ud}V_{ub}|$). This leads to tree-level contributions that are much larger than the penguin contributions, resulting in a direct $CP$ asymmetry of approximately $18.4\%$. For $B_s\to \bar \kappa^0\pi^0$, the tree-level contributions are suppressed by the small Wilson coefficient $C_1+C_2/3$, and thus the direct $CP$ asymmetry reaches about $37\%$. In the cases of $B_s\to a_0^+(980)K^-$ and $B_s\to a_0^0(980)\overline{K}^0$, the tree-level emission-type contributions are highly suppressed due to the small vector decay constant of $a_0^+(980)$, or they vanish entirely because $f_s=0$ for $a_0^0(980)$. As a result, the small penguin contributions become comparable to the suppressed tree-level contributions, leading to large direct $CP$ asymmetries of $98\%$ and $72\%$, respectively. We recommend that the experimentalists focus on measuring the $B_s\to a_0(980)K$ decays, as these processes exhibit both significant $CP$ violation and large branching fractions, making them well-suited for experimental measurement.

Although a consistent picture exists \cite{Close:2002zu}, supported by some experimental data, suggesting that scalar mesons above 1 GeV can be described as a conventional $q\bar{q}$ nonet with possible glueball admixture, mesons below 1 GeV are typically modeled as a $qq\bar{q}\bar{q}$ nonet, with potential interference from $0^+$, $q\bar{q}$, and glueball states \cite{Jaffe:1976ig, Jaffe:1976ih}. The strong couplings of $f_0(980)$ and $a_0(980)$ to the $K\overline{K}$ pair, the broader decay widths of $f_0(500)$ and $\kappa$ compared to $f_0(980)$ and $a_0(980)$, and the fact that $a_0(980)$ is heavier than $\kappa$, all suggest a more complicated picture. In the four-quark model, additional non-factorizable diagrams arise, where one pair of $q\bar{q}$ in the scalar meson is produced via a soft gluon, making these contributions more intricate and challenging to calculate. Furthermore, the calculations of decay constants for scalar mesons and the $B_s \to S$ form factors within the four-quark model goes beyond the scope of the conventional quark model. Intuitively, it is expected that the distribution amplitude in the four-quark model would be smaller compared to the two-quark picture \cite{Cheng:2005nb}.

\begin{figure}[!htb]
    \centering
    \includegraphics[width=0.45\linewidth]{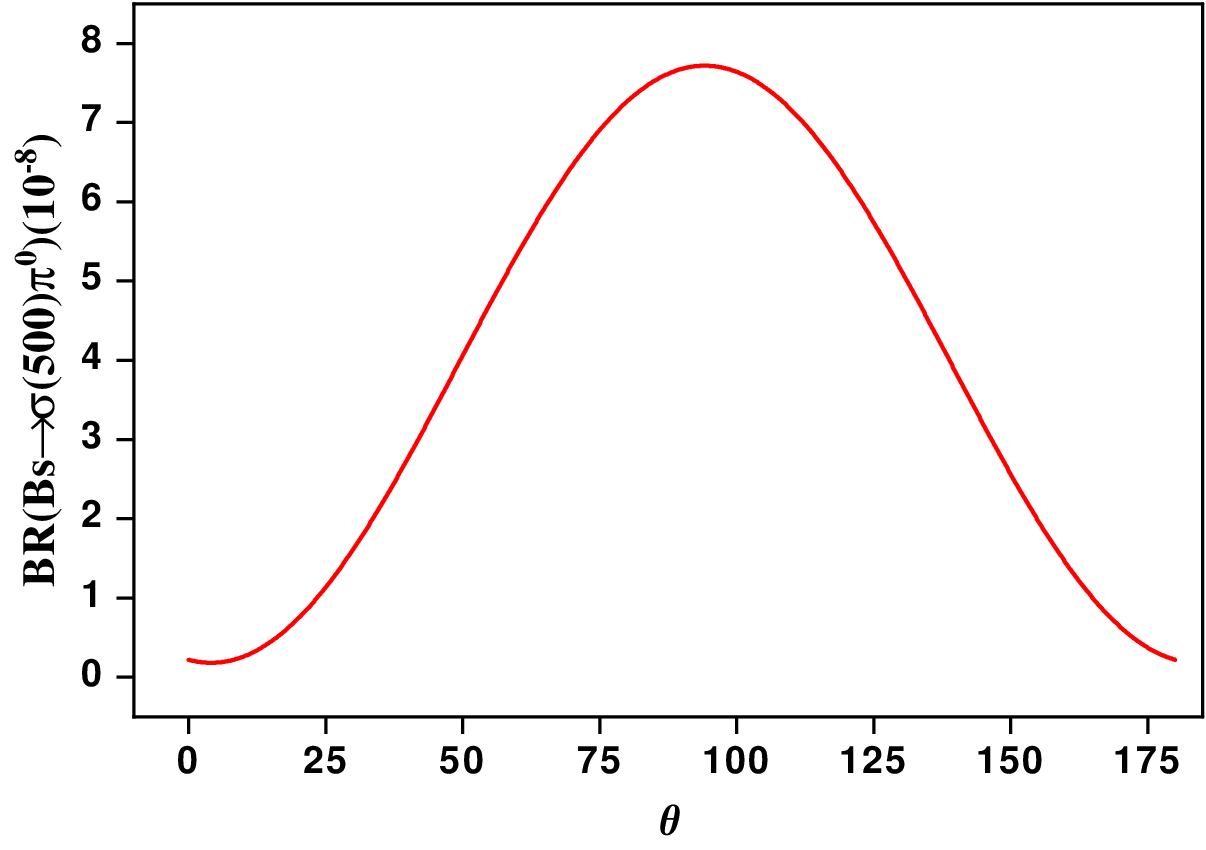}    
    \includegraphics[width=0.45\linewidth]{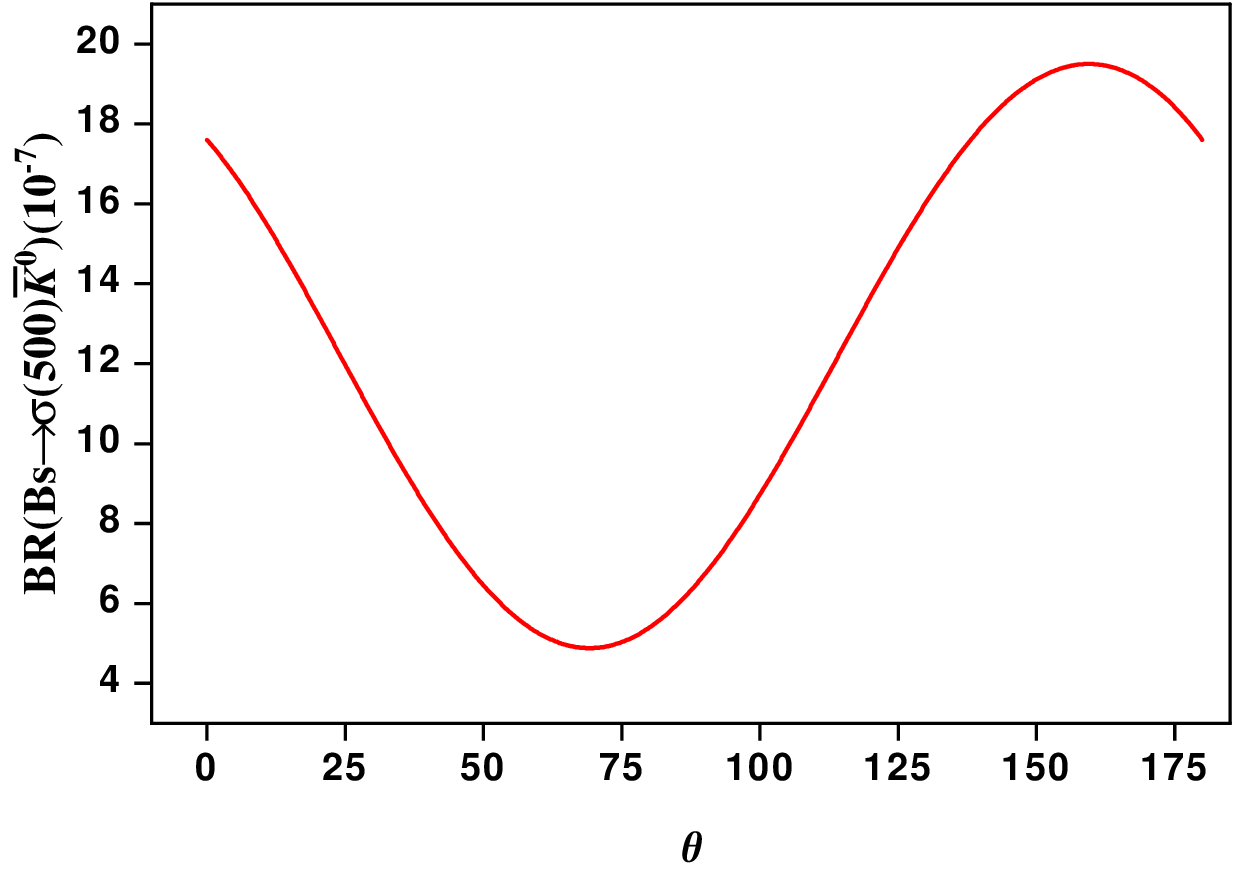}   
    \includegraphics[width=0.45\linewidth]{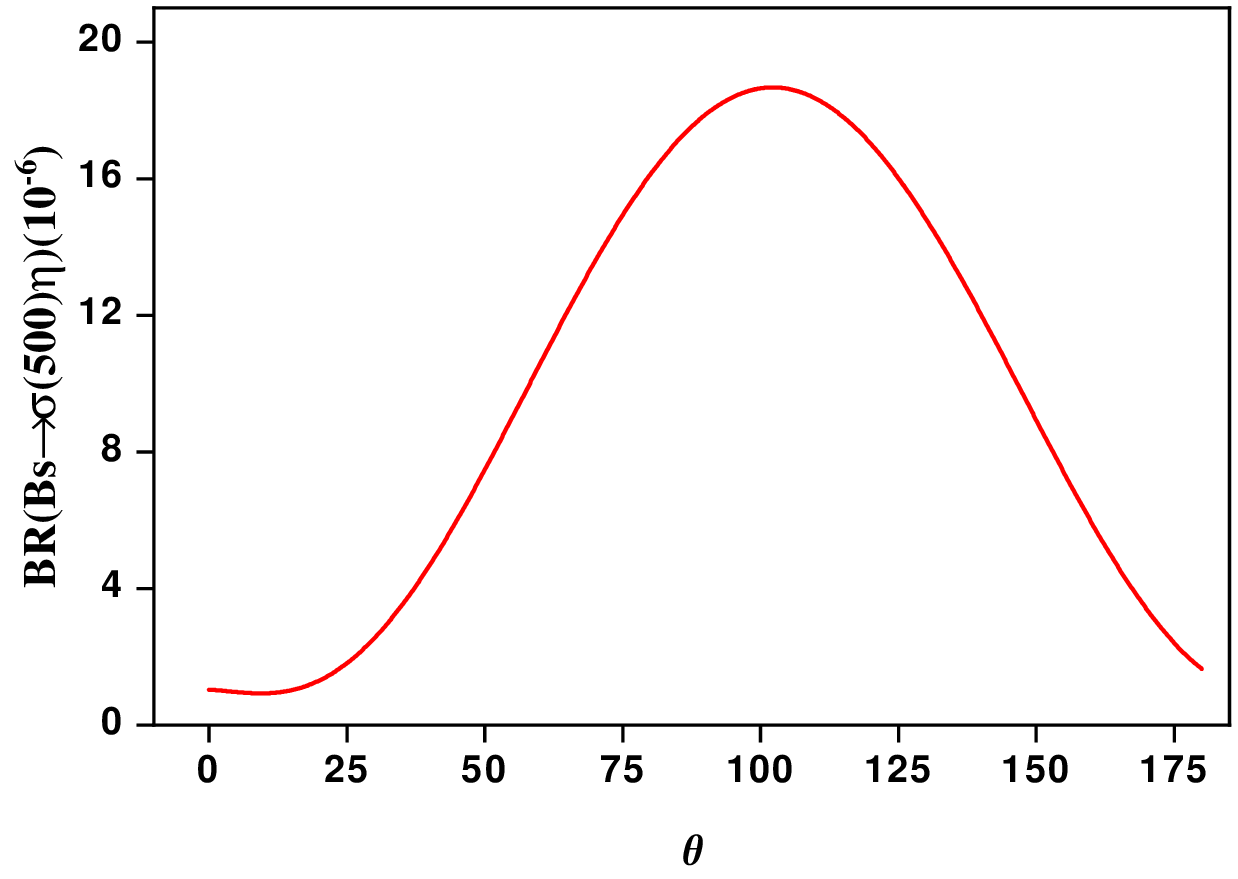}   
    \includegraphics[width=0.45\linewidth]{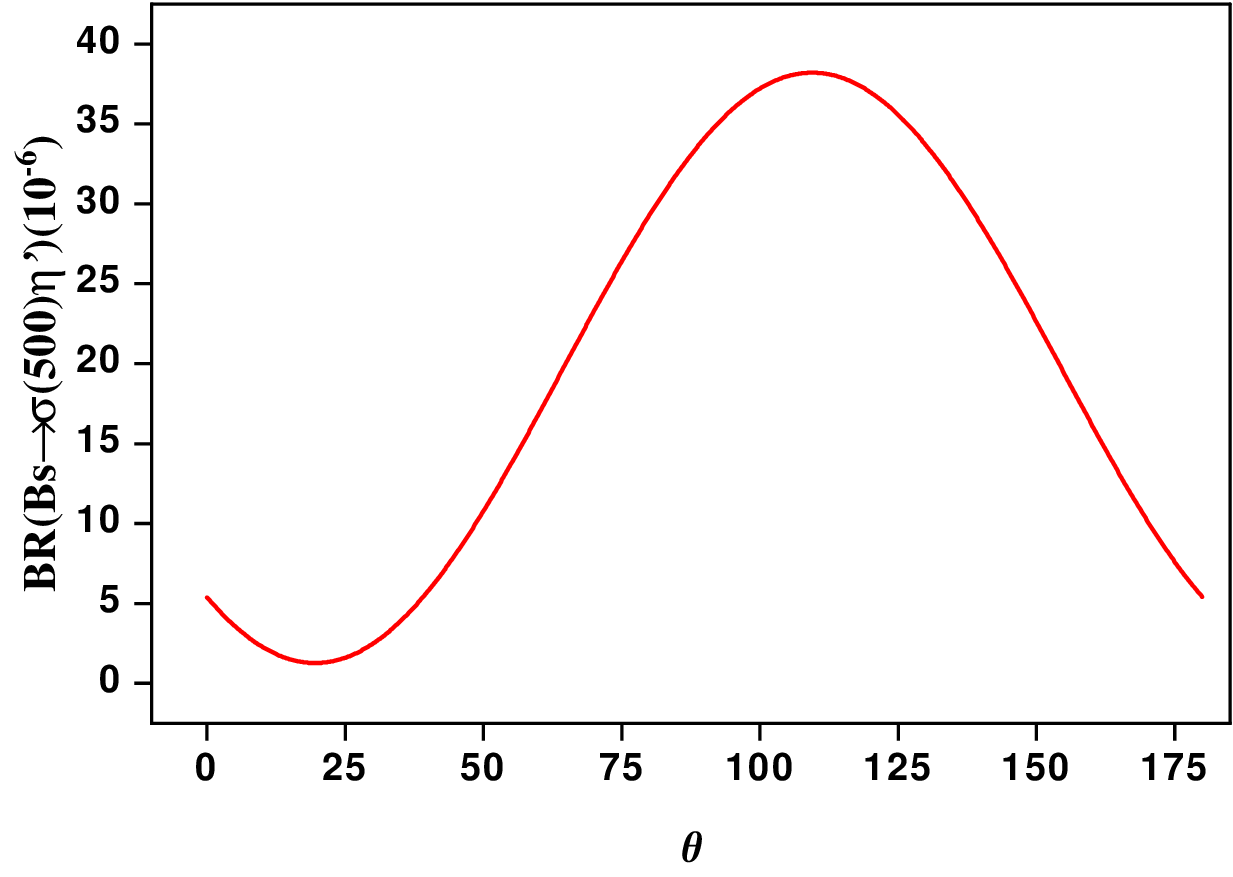} 
    \includegraphics[width=0.45\linewidth]{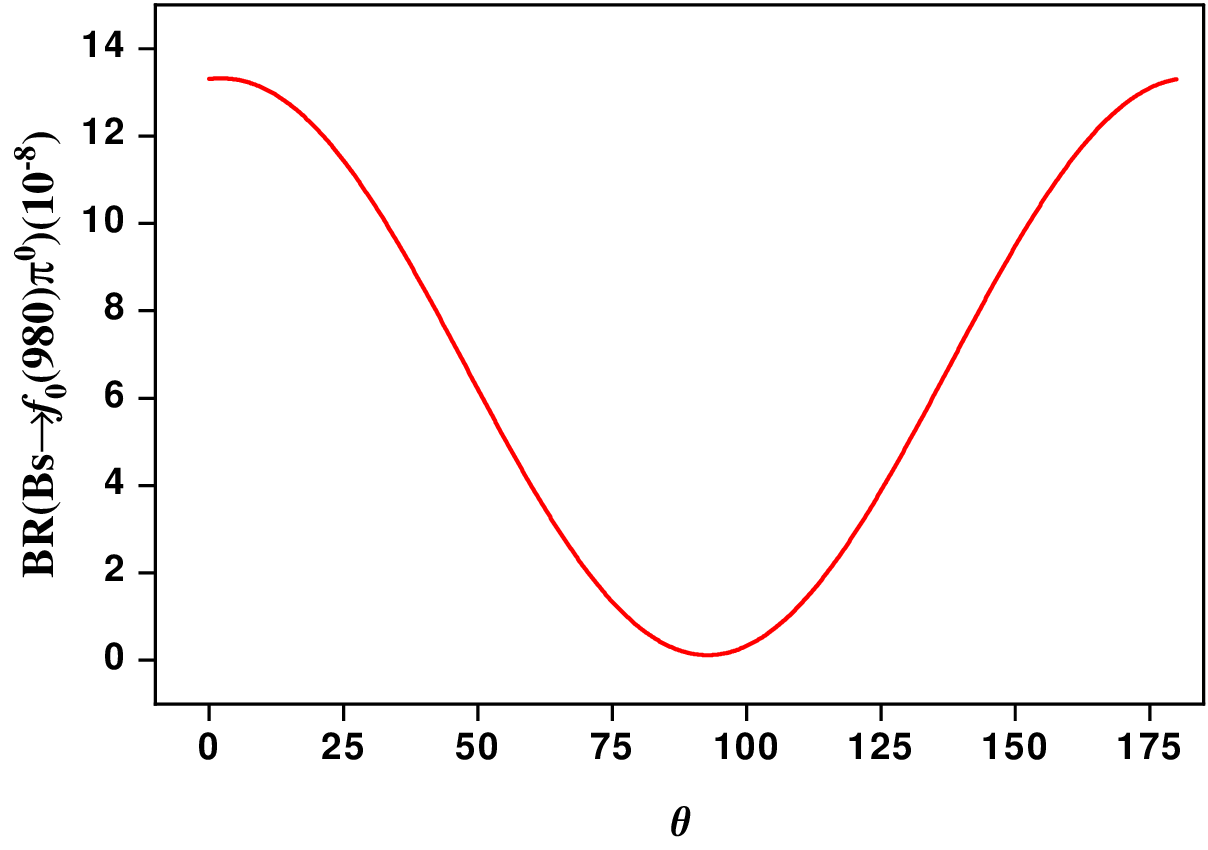}    
    \includegraphics[width=0.45\linewidth]{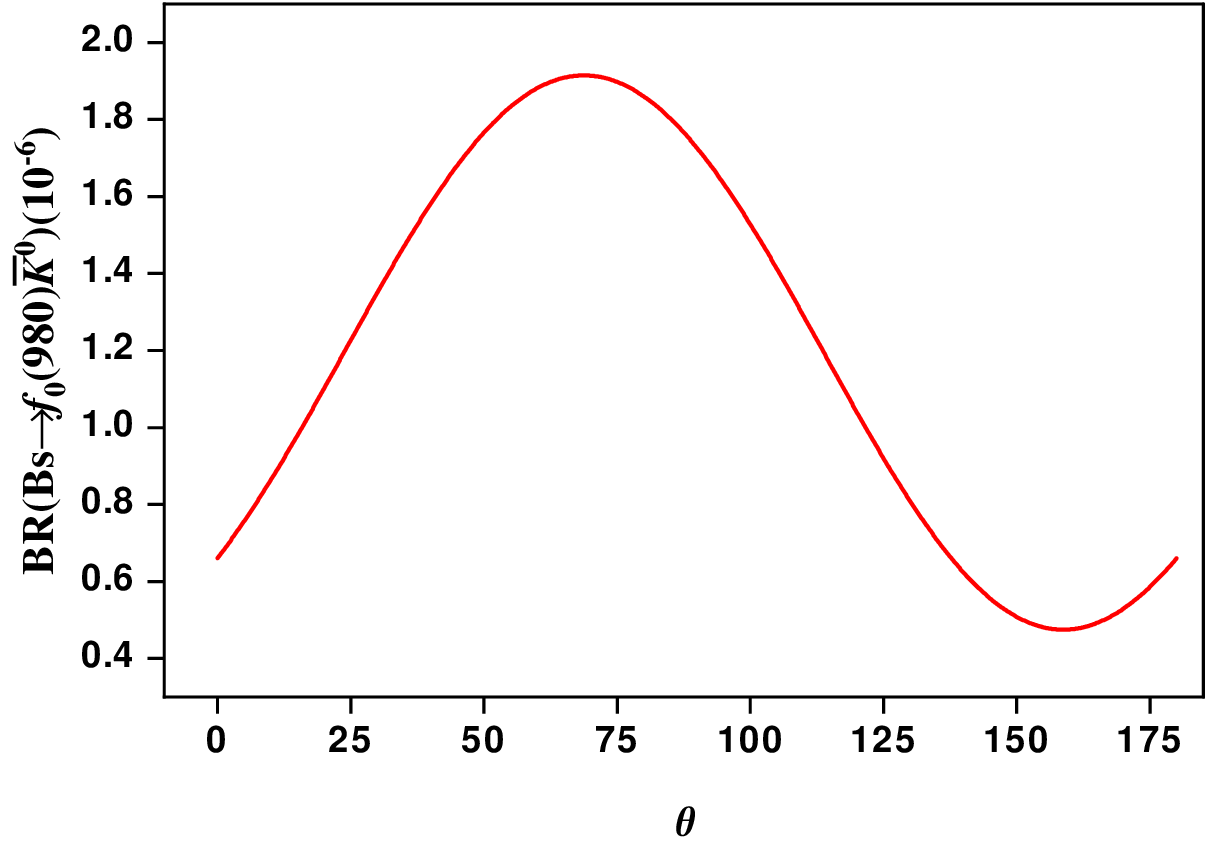}   
    \includegraphics[width=0.45\linewidth]{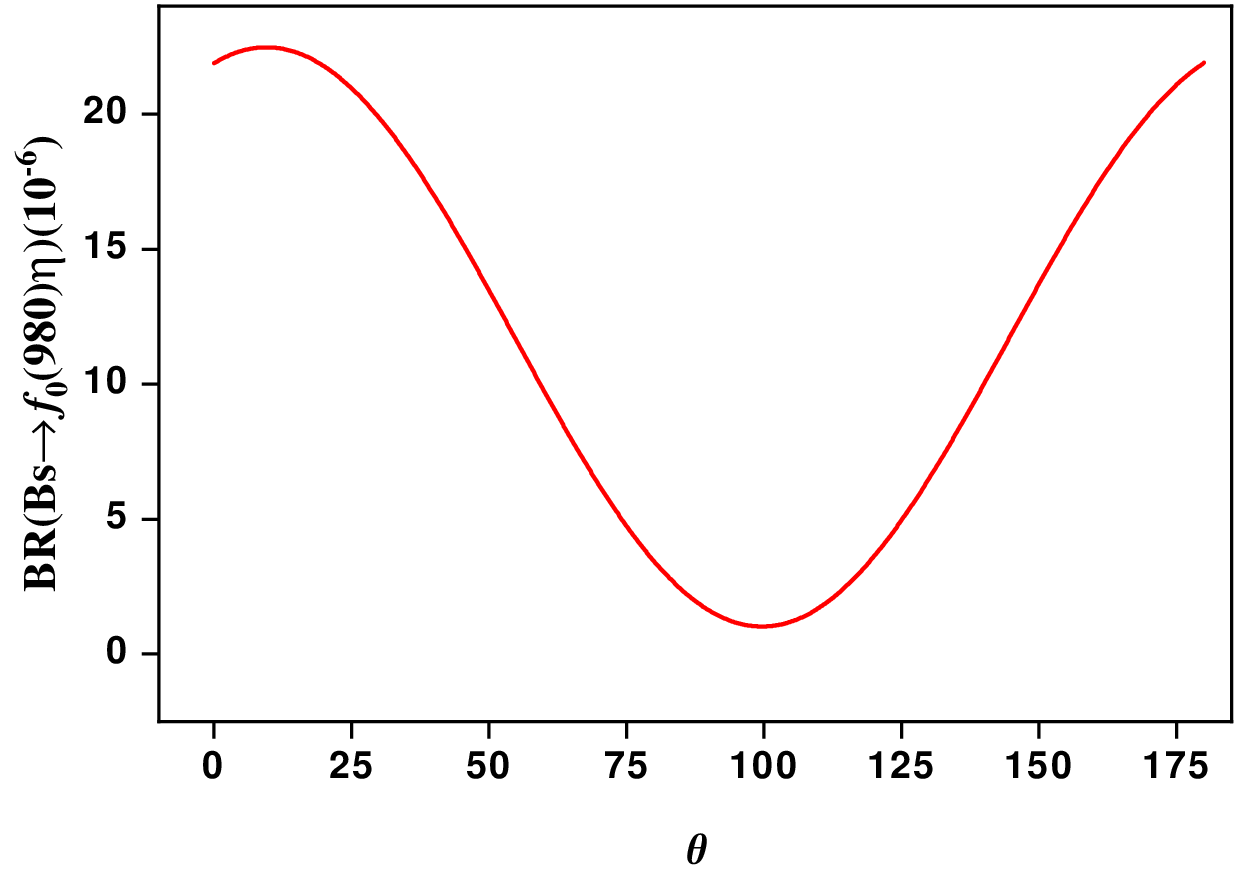}   
    \includegraphics[width=0.45\linewidth]{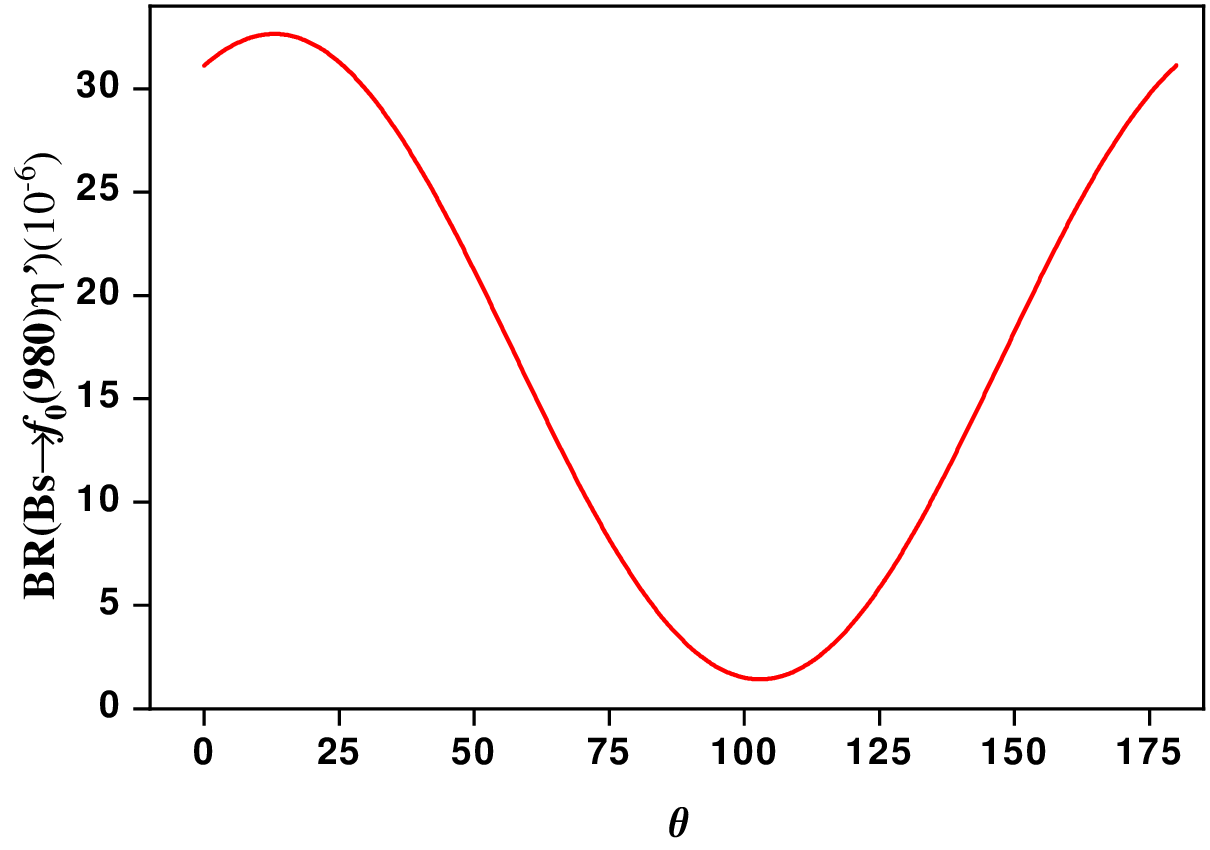} 
     \caption{The branching fractions of the $B_{s}\to f_{0}(980)[\sigma]P$ decays versus the $f_{0}(980)-\sigma$ mixing angle $\theta$.}
    \label{fig:BR}
\end{figure}

\begin{figure}[!htb]
    \centering
    \includegraphics[width=0.45\linewidth]{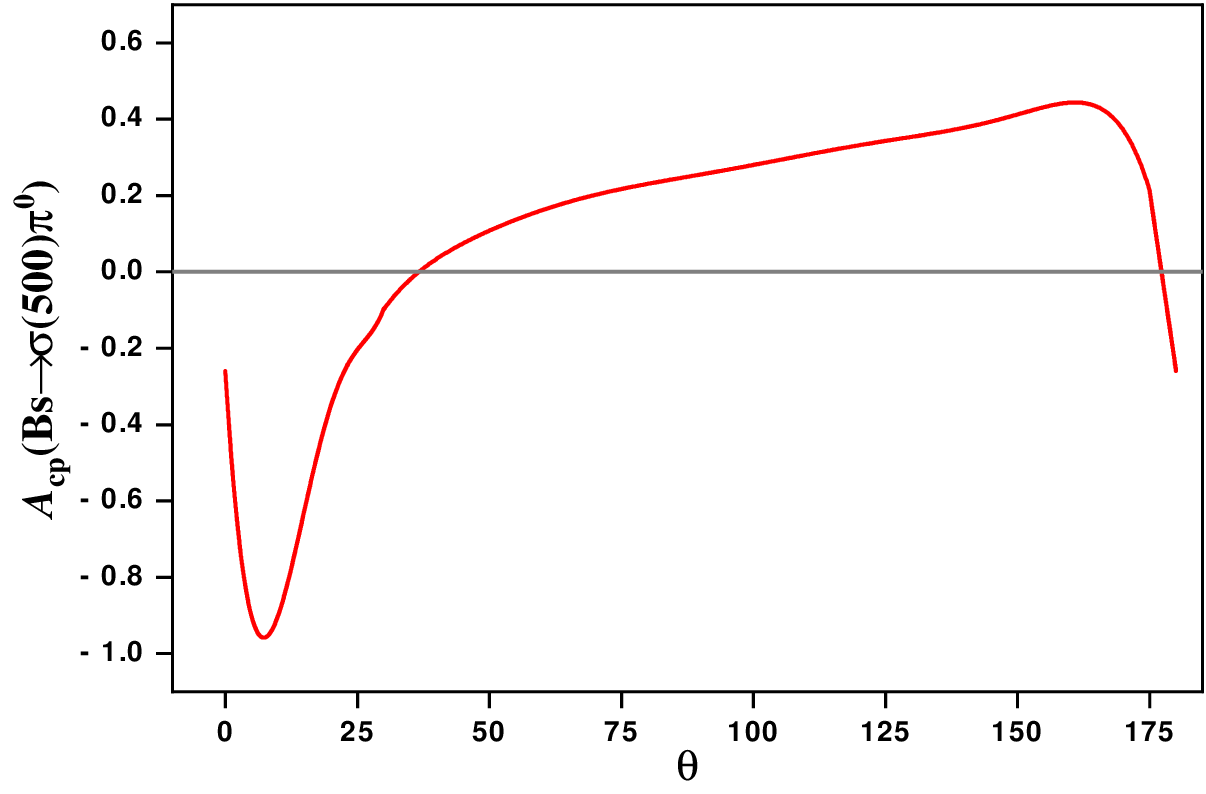}    
    \includegraphics[width=0.45\linewidth]{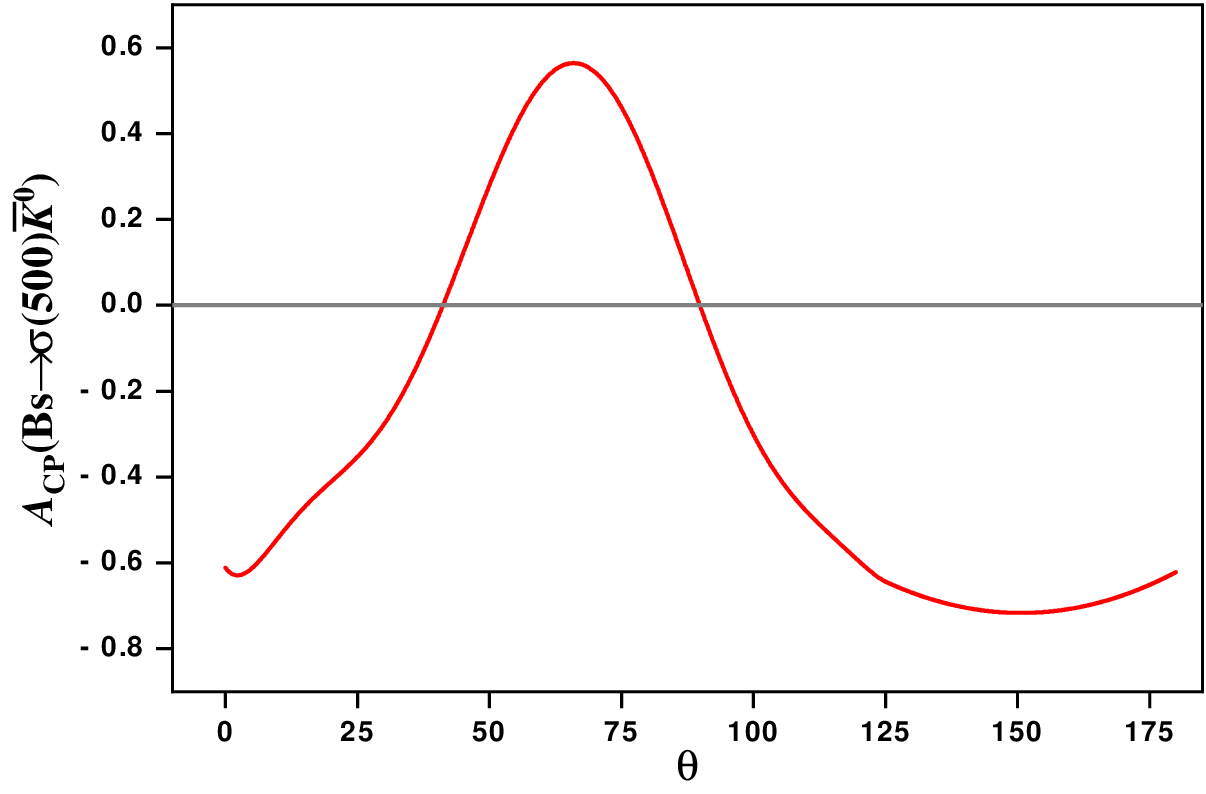}   
    \includegraphics[width=0.45\linewidth]{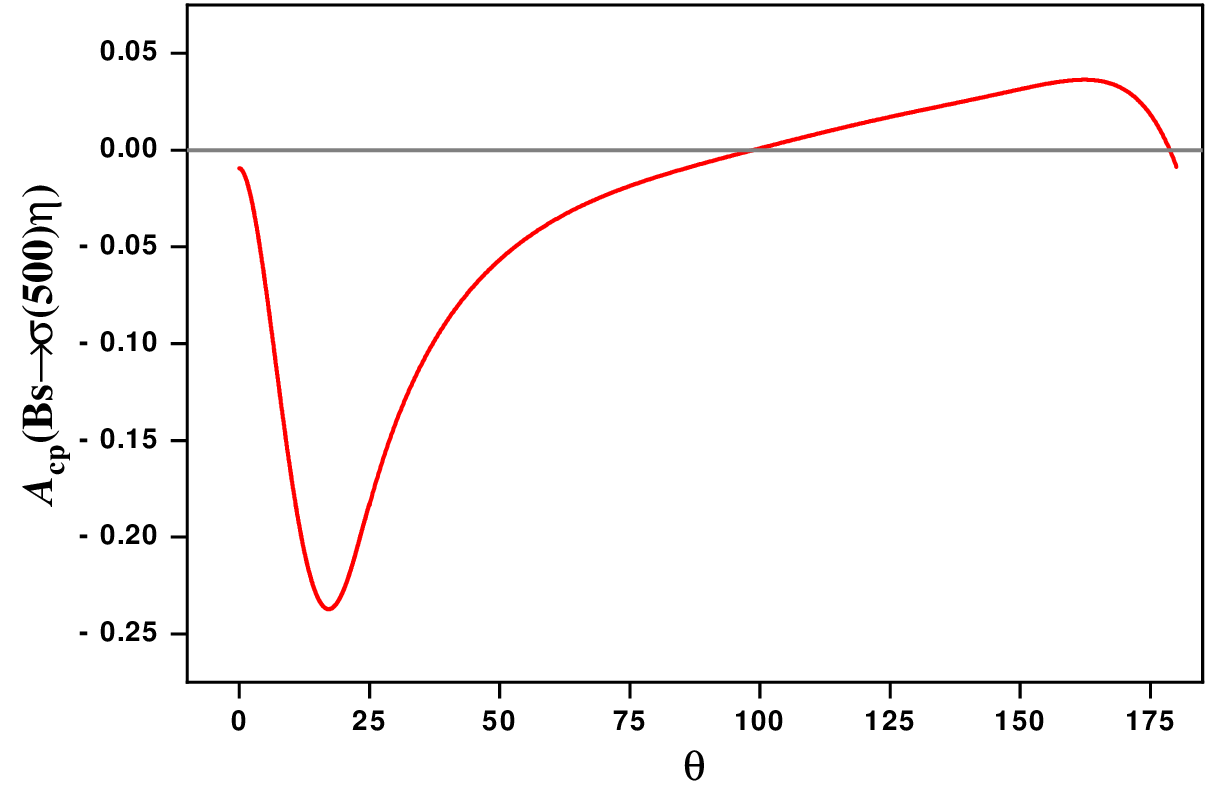}   
    \includegraphics[width=0.45\linewidth]{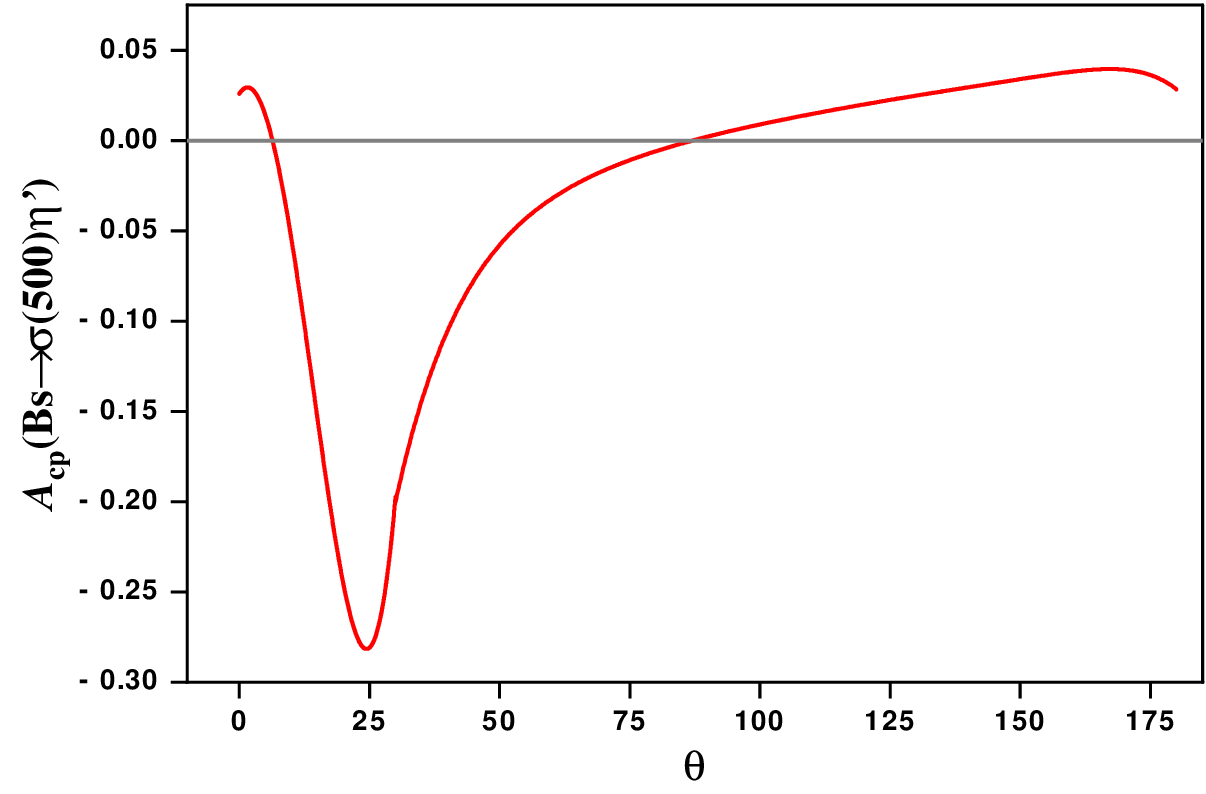} 
    \includegraphics[width=0.45\linewidth]{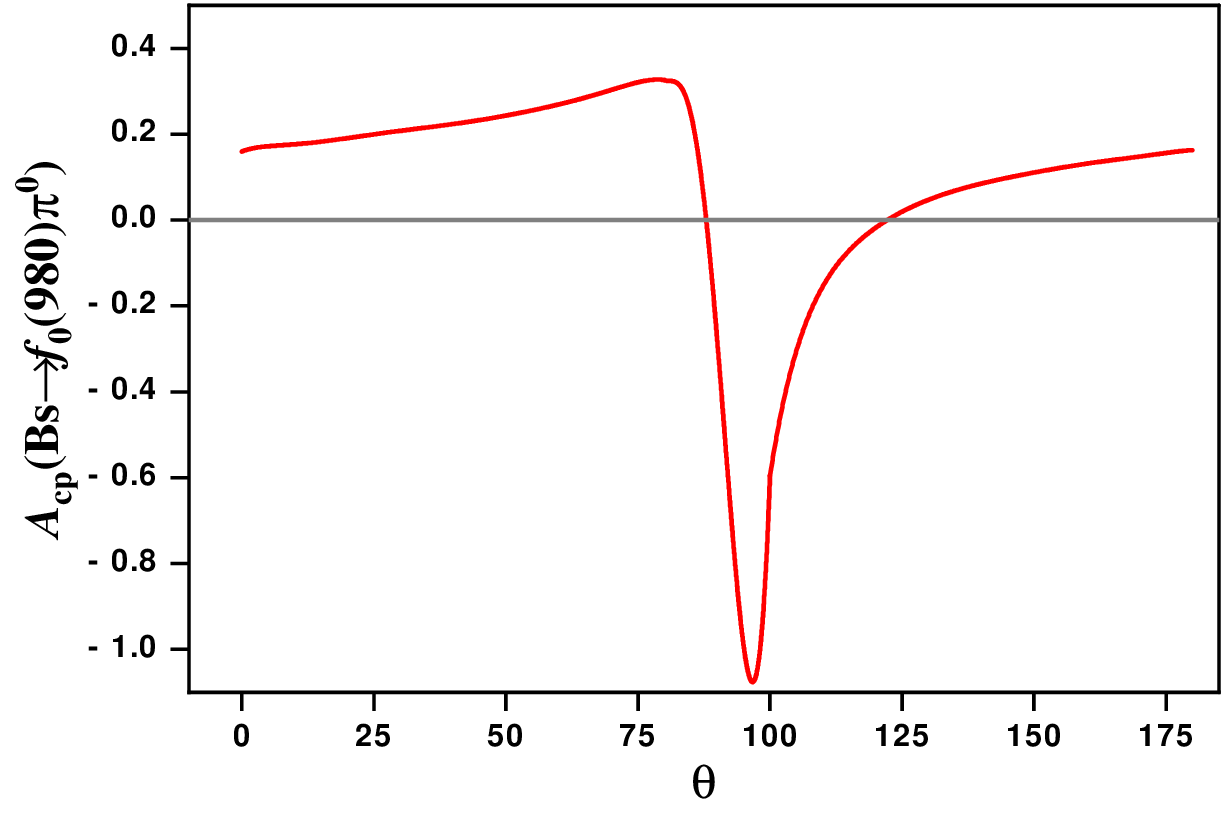}    
    \includegraphics[width=0.45\linewidth]{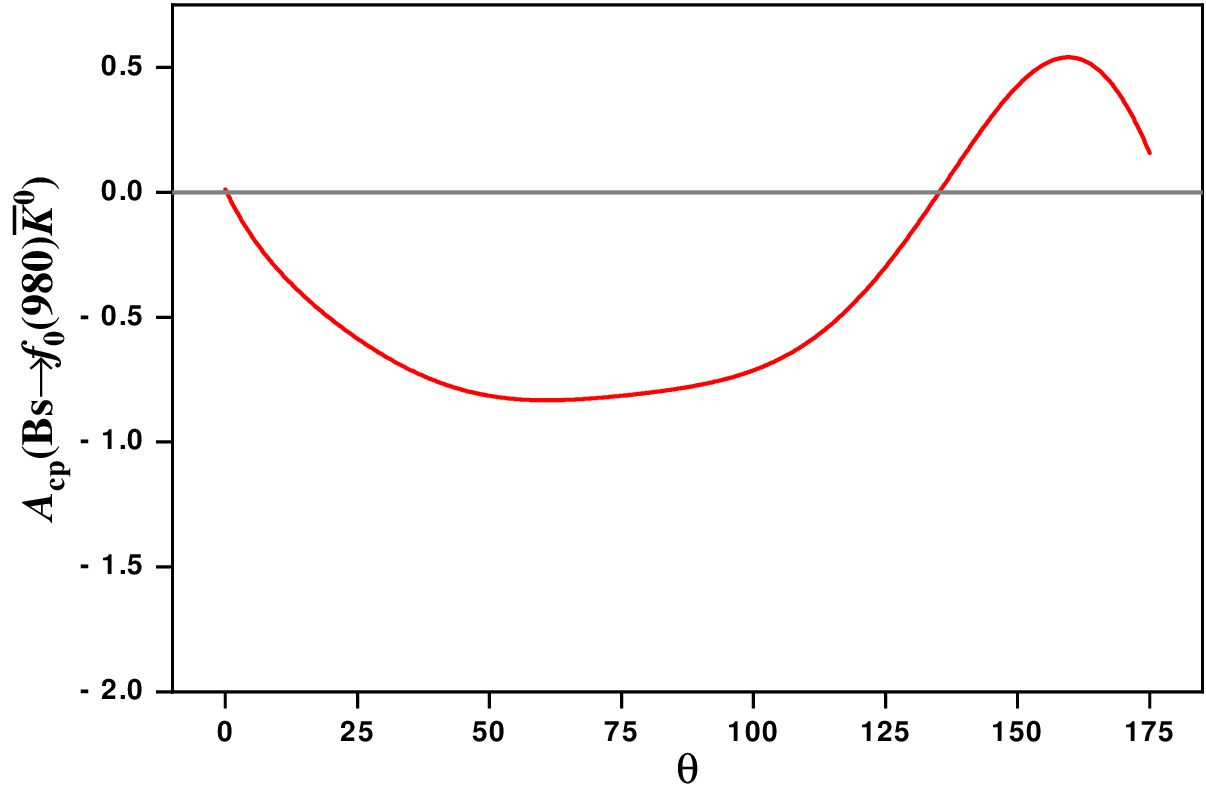}   
    \includegraphics[width=0.45\linewidth]{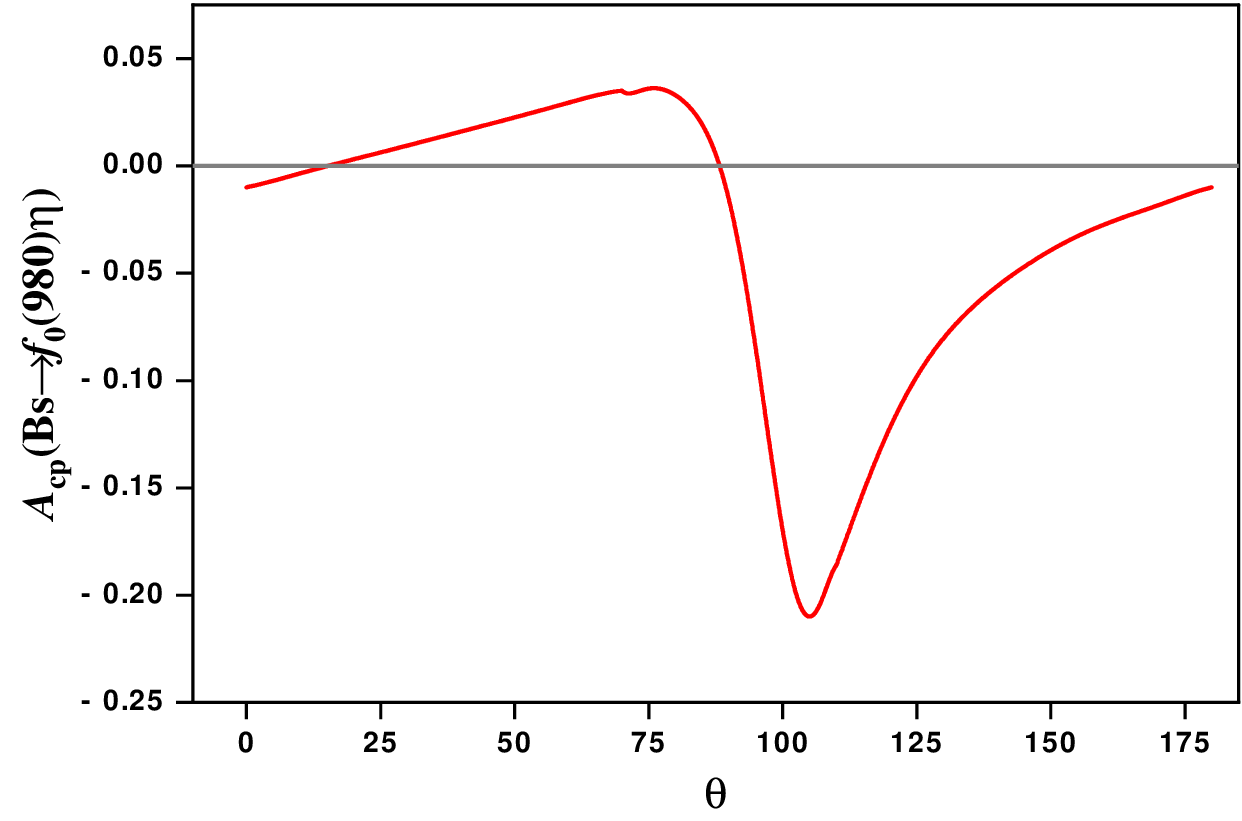}   
    \includegraphics[width=0.45\linewidth]{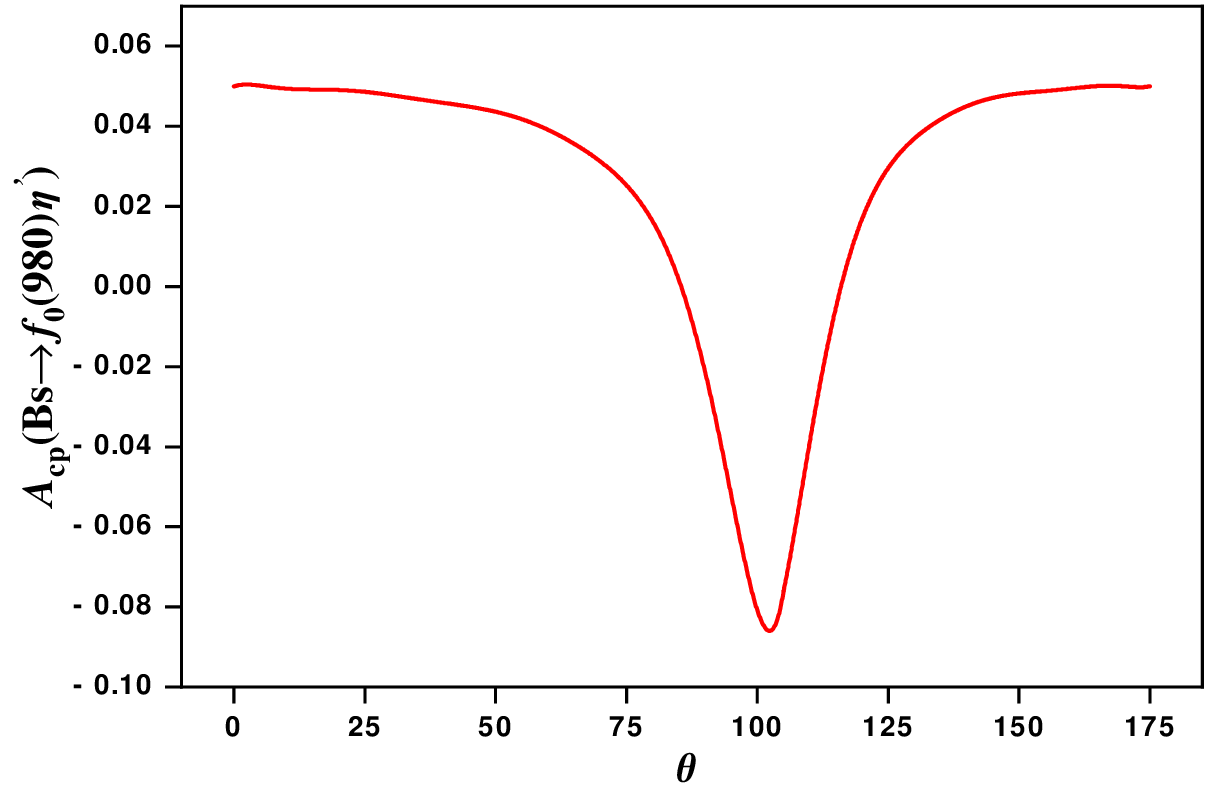} 
    \caption{The direct $CP$ asymmetries of the $B_{s}\to f_{0}(980)[\sigma]P$ decays versus the $f_{0}(980)-\sigma$ mixing angle $\theta$.}
    \label{fig:CP}
\end{figure}

In the conventional two-quark picture and the ideal mixing model for $f_0(980)$ and $\sigma(500)$,  $f_0(980)$ is  purely an $s\bar{s}$ state, while $\sigma(500)$ is viewed as a $n\bar{n}$ state, with $n\bar{n}=(u\bar{u}+d\bar{d})/\sqrt{2}$, as suggested by experimental data from  $D_s^+\to f_0(980)\pi^+$ and $\phi\to f_0(980) \gamma$ \cite{ParticleDataGroup:2024cfk}. However, experimental measurements such as $\Gamma(J/\psi\to f_0(980)\omega)\approx\frac{1}{2}\Gamma(J/\psi\to f_0(980)\phi)$ \cite{ParticleDataGroup:2024cfk} clearly indicate that the $f_0(980)$ contains a significant $n\bar{n}$ component. This is further supported by the fact that the decay width of $f_0(980)$ is similar to that of $a_0(980)$ and is dominated by $\pi\pi$ processes. Thus, the isoscalars  $\sigma(500)$ and $f_0(980)$ mix according to the following relations:
\begin{eqnarray}
f_0(980)&=&s\bar{s}\cos\theta+n\bar{n}\sin\theta,\\
\sigma(500)&=&-s\bar{s}\sin\theta+n\bar{n}\cos\theta.
\end{eqnarray}
The mixing angle $\theta$ of the $\sigma(500)-f_0(980)$ system has been extensively studied in various experimental contexts, with detailed results summarized in Refs.~\cite{Alford:2000mm, Anisovich:2001zp, Gokalp:2004ny}. Two commonly accepted ranges for $\theta$ are $25^\circ < \theta < 40^\circ$ and $140^\circ < \theta < 165^\circ$, which are frequently used in phenomenological analyses. Using these two widely adopted ranges for the mixing angle, we calculate the branching fractions and direct $CP$ asymmetries for the decays $B_s \to f_0(980)/\sigma(500) P$, with the results presented in Table.~\ref{table:2}. From the table, we observe that the branching fraction for $B_s \to \sigma \eta^{\prime}$ shows a significant difference between the two ranges of $\theta$. Once experimental data become available, these results will be useful in determining whether the mixing angle is acute or obtuse. For the other decays, the differences in both branching fractions and direct $CP$ asymmetries between the two angle ranges are not large enough to conclusively determine the mixing angle. Additionally, we present the dependence of the branching fractions and direct $CP$ asymmetries on the mixing angle $\theta$ in Figs.~\ref{fig:BR} and \ref{fig:CP}, respectively. These curves are expected to provide valuable guidance in determining the mixing angle when combined with ongoing experimental measurements. 


In Ref.~\cite{LHCb:2019vww}, the LHCb collaboration analyzed  the first untagged decay-time-integrated amplitude of $B_s\to K_s^0 K^{\pm}\pi^{\mp}$ decays. Based on this analysis, employing the narrow-width approximation, the branching fractions for the corresponding two-body decays $B_s\to K_0^{*\pm}(1430)K^{\mp}$ and $B_s\to K_0^{*0}(1430)\overline{K}^0/\overline{K}_0^{*0}(1430)K^0$ were obtained as
\begin{eqnarray}
{\cal B}(B_s\to K_0^{*\pm}(1430)K^{\mp})&=&(31.3\pm2.3\pm0.7\pm25.1\pm3.3)\times10^{-6},\\
{\cal B}(B_s\to K_0^{*0}(1430)\overline{K}^0/\overline{K}_0^{*0}(1430)K^0)&=&(33.0\pm2.5\pm0.9\pm9.1\pm3.5)\times10^{-6}.
\end{eqnarray}
In this work, we calculate the branching fractions for these decays using PQCD approach for the first time. Under two different scenarios, our results are given as follows:
\begin{align}
\begin{split}
{\cal B}(B_s\to K_0^{*\pm}(1430)K^{\mp})= \left \{
\begin{array}{ll}
(43.6^{+15.7+10.8+3.5}_{-12.5-8.7-1.1})\times10^{-6} \;\;&{\rm scenario~1},\\
(54.9^{+22.9+14.1+2.0}_{-19.1-10.6-2.3})\times10^{-6} \;\;& {\rm scenario~2};
\end{array}
\right.
\end{split}
\end{align}
\begin{align}
\begin{split}
{\cal B}(B_s\to K_0^{*0}(1430)\bar{K}^0/\bar{K}_0^{*0}(1430)K^0)= \left \{
\begin{array}{ll}
(43.4^{+14.6+11.3+0.7}_{-14.2-11.8-2.1})\times10^{-6}\;\;&{\rm scenario~1},\\
(54.0^{+26.7+7.6+1.0}_{-21.6-10.6-2.5})\times10^{-6}\;\;& {\rm scenario~2}.
\end{array}
\right.
\end{split}
\end{align}
From the above results, it is clear that both the experimental measurements and theoretical predictions have large uncertainties. Considering these uncertainties, our theoretical calculations are consistent with the experimental measurements. In Ref.~\cite{Chen:2021oul}, these processes had been also evaluated in QCDF, and the obtained branching fractions are presented as:
\begin{align}
\begin{split}
{\cal B}(B_s\to K_0^{*\pm}(1430)K^{\mp})= \left \{
\begin{array}{ll}
(7.19^{+0.33+0.89+4.38}_{-0.24-1.51-3.38})\times10^{-6},\;\;&{\rm scenario~1}\\
(33.73^{+1.51+4.17+18.13}_{-1.08-7.09-14.57})\times10^{-6},\;\;& {\rm scenario~2}
\end{array}
\right.
\end{split}
\end{align}
\begin{align}
\begin{split}
{\cal B}(B_s\to K_0^{*0}(1430)\bar{K}^0/\bar{K}_0^{*0}(1430)K^0)= \left \{
\begin{array}{ll}
(6.62^{+0.29+0.88+4.37}_{-0.21-1.48-3.35})\times10^{-6},\;\;&{\rm scenario~1}\\
(32.78^{+1.45+4.18+18.02}_{-1.03-7.09-14.49})\times10^{-6}.\;\;& {\rm scenario~2}
\end{array}
\right.
\end{split}
\end{align}
It is evident that our results show significant deviations from those obtained in \cite{Chen:2021oul} using QCDF. These discrepancies can be attributed to two main factors: first, differences in the values of non-perturbative parameters, particularly the LCADs of  $K_0^*(1430)$; and second, the handling of unavoidable endpoint singularities, which are parameterized by mode-dependent parameters ($\rho_{H, A}$ and $\phi_{H, A}$) in QCDF and smoothed by keeping the intrinsic transverse momenta in PQCD.  In this regard, we call for high-precision non-perturbative LCDAs to improve the accuracy of theoretical calculations. On the other hand, more advanced experimental data processing methods are need to reduce statistical and systematic uncertainties in experimental results. Only by combining both can we better test the QCDF and PQCD methods and further explore the properties and internal structure of $K_0^*(1430)$.

Tables.~\ref{table:3} and \ref{table:4} summarize the branching fractions and direct $CP$ asymmetries for the $B_s\to SP$ decays involving heavier scalar mesons, considering two different scenarios. It is evident that the branching  fractions for penguin-dominated decay channels are generally larger than those for tree-dominated processes. This is because, in the tree-dominated decays, the tree operator amplitudes are either suppressed or vanish due to the small vector decay constants of the scalar mesons, while the contributions from the penguin operators are further reduced by the small CKM matrix elements $V_{tb}V_{td}$. Among the tree-dominated decays, the branching fraction for $B_s\to \overline{K}_0^{*-}(1430)\pi^+$ is slightly larger than those of the other tree-dominated processes. This is because the $\pi^+$ meson is emitted in this channel, allowing the tree-type amplitude to remain unsuppressed. In contrast, for the decay $B_s\to \overline{K}_0^{*0}(1430)\pi^0$, the tree-type contribution is suppressed by the small Wilson coefficients $C_1+C_2/3$, which results in a much smaller branching fraction compared to $B_s\to \overline{K}_0^{*-}(1430)\pi^+$. Additionally, the direct $CP$ asymmetries for penguin-dominated decays are generally smaller than those for tree-dominated decays. In the penguin-dominated decays, the tree contributions are heavily suppressed due to both the small CKM elements and the vector decay constants, making them much smaller than the penguin contributions. As a result, the direct $CP$ asymmetries are reduced. Conversely, in tree-dominated decays, the tree contributions, while suppressed by small vector decay constants and CKM elements, are still comparable to the penguin contributions. Hence, the direct $CP$ asymmetries in tree-dominated decays tend to be large, as these asymmetries are proportional to the interference between the tree and penguin contributions. For instance, the direct $CP$ asymmetry for the tree-dominated decay $B_s\to K_{0}^{*0}(1430)\pi^0$ reaches $95\%$ in scenario 1 and $83\%$ in scenario 2, due to the suppression of the tree contribution by the small Wilson coefficient $C_1+C_2/3$.

In the tables, we provide a comparison of our results with earlier PQCD predictions from Refs.~\cite{Zhang:2016qvq, Zhang:2013efa, Zhang:2012zze, Zhang:2010kw, Liu:2009xm}. Our findings are generally consistent with these previous predictions, with any observed discrepancies attributable to the improvements introduced in this work, as discussed earlier. For the majority of decay channels, the differences in branching fractions between the two scenarios are not significant enough to definitively determine which scenario is more appropriate. However, for the decay $B_s\to f_0(1370)\eta$, a notable discrepancy emerges, with substantial differences in the branching fractions between the two scenarios, given by: 
\begin{align}
{\cal B}(B_s \to f_0(1370)\eta)= \left \{
\begin{array}{ll}
(1.25^{+0.69+0.64+0.05}_{-0.54-0.48-0.08}\times10^{-6}\;\;&{\rm scenario~1},\\
(14.3^{+9.0+3.1+0.4}_{-6.4-2.8-0.3})\times10^{-6}\;\;& {\rm scenario~2}.
\end{array}
\right.
\end{align}
This substantial variation provides a clear distinction for identifying the more favored scenario. Additionally, the direct $CP$ asymmetries for the decays $B_s\to a_0^+(1450)K^-$ and $B_s \to a_0^0(1450)\overline {K}^0$  are highly sensitive to the choice of scenario. In scenario 1, the direct $CP$ asymmetries are $3.28\%$ and $-7.47\%$, respectively. In contrast, in scenario 2, these asymmetries increase dramatically to $72.5\%$ and $94.2\%$. Future high-precision measurements of these observables could help differentiate between the two scenarios and provide insights into the internal structure of scalar hadrons.
\section{Summary}\label{sec:4}
In this study, we investigate the $B_s\to SP$ decays within the PQCD framework to predict their branching fractions and $CP$ asymmetries.  Our results align well with prior PQCD predictions and experimental measurements from LHCb, with any discrepancies remaining within expected uncertainty ranges. We observe that the branching fractions for penguin-dominated $B_s\to SP$  decays are significantly larger than those for tree-dominated decays. This disparity arises because the tree-level contributions either vanish or are strongly suppressed by the small vector decay constants of the scalar mesons. Consequently, the direct $CP$  asymmetries for tree-dominated decays tend to be relatively large, as the penguin contributions—suppressed by $|V_{td}V_{td}|$ —are of comparable magnitude to the already suppressed tree-level contributions. Conversely, direct $CP$ asymmetries for penguin-dominated decays are generally small, as the tree-level contributions are doubly suppressed by both the small CKM elements and the vector decay constants of the scalar mesons. Furthermore, certain decays display pronounced sensitivity to the choice of theoretical scenario, with both branching fractions and CP asymmetries serving as discriminators between two scenarios. These results underscore the potential of $B_s \to SP$ decays as a probe for understanding the nature of scalar mesons and testing PQCD approach. 

\section*{Acknowledgment}
This work was supported by National Natural Science Foundation of China under Grants Nos.12375089 and 12435004, and by the Natural Science Foundation of Shandong province under the Grant No. ZR2022MA035 and ZR2022ZD26.
\bibliographystyle{bibstyle}
\bibliography{refs}
\end{document}